\documentclass[10pt,journal,compsoc]{IEEEtran}
\usepackage{cite}
\usepackage{tikz}
\usepackage{amsmath}
\usepackage{amsfonts}
\usepackage[ruled,vlined,linesnumbered]{algorithm2e}
\usepackage[noend]{algpseudocode}
\usepackage{subcaption}
\usepackage{booktabs}
\usepackage{multirow}
\usepackage{xcolor}
\usepackage{comment}
\usepackage{multicol}

\SetCommentSty{mycommfont}
\usepackage{ragged2e}
\usepackage{colortbl}
\usepackage{array}

\newcolumntype{P}[1]{>{\centering\arraybackslash}p{#1}}
\newcolumntype{M}[1]{>{\centering\arraybackslash}m{#1}}

\newcommand*\rot{\rotatebox{90}}
\newcommand*\rotn{\rotatebox{90}}

\newcommand{\name}{{\sf SUNDEW }}
\newcommand{\nameA}{{\sf SUNDEW}}
\usepackage{tikz}
\usetikzlibrary{backgrounds,calc,patterns}
\usetikzlibrary{shapes,arrows,fit}
\tikzstyle{block} = [draw, rectangle, text width=2cm, text centered, minimum height=1.2cm, node distance=3cm,fill=white]
\tikzstyle{container} = [draw, rectangle, inner sep=0.3cm, fill=gray,minimum height=3cm]

\tikzstyle{blockincontainer} = [draw, rectangle, text width=2cm, text centered, fill=white,  minimum size=2cm,
   anchor=south west]

\tikzset{container/.style={
   draw,
   minimum size=2cm,
  anchor=south west
 },
 bg box/.style={
   container,
   anchor=center
 }
}

\begin{document}

\title{\nameA: An Ensemble of Predictors for Case-Sensitive Detection of Malware}

\author{Sareena~Karapoola,
        Nikhliesh~Singh,
        Chester~Rebeiro,
        and~Kamakoti~V
\IEEEcompsocitemizethanks{\IEEEcompsocthanksitem S. Karapoola, N. Singh, C. Rebeiro and Kamakoti V. are with the Department of
CSE, Indian Institute of Technology Madras, Chennai, India.
\protect\\

E-mail: \{sareena, nik, chester, kama\}@cse.iitm.ac.in.}
}

\markboth{}%
{Shell \MakeLowercase{\textit{et al.}}: Bare Demo of IEEEtran.cls for Computer Society Journals}

\IEEEtitleabstractindextext{%

\begin{abstract}
\justifying Malware programs are diverse, with varying objectives, functionalities, and threat levels ranging from mere pop-ups to financial losses. Consequently, their run-time footprints across the system differ, impacting the optimal data source (Network, Operating system (OS), Hardware) and features that are instrumental to malware detection. Further, the variations in threat levels of malware classes affect the user requirements for detection. Thus, the optimal tuple of $\langle \tt data$-$\tt source$, $\tt features$, $\tt user$-$\tt requirements \rangle$ is different for each malware class, impacting the state-of-the-art detection solutions that are agnostic to these subtle differences. 

This paper presents \nameA, a framework to detect malware classes using their optimal tuple of $\langle \tt data$-$\tt source$, $\tt features$, $\tt user$-$\tt requirements \rangle$. \name uses an ensemble of specialized predictors, each trained with a particular data source (network, OS, and hardware) and tuned for features and requirements of a specific class. While the specialized ensemble with a holistic view across the system improves detection, aggregating the independent conflicting inferences from the different predictors is challenging. \name resolves such conflicts with a hierarchical aggregation considering the threat-level, noise in the data sources, and prior domain knowledge. We evaluate \name on a real-world dataset of over 10,000 malware samples from 8 classes. It achieves an F1-Score of one for most classes, with an average of 0.93 and a limited performance overhead of $1.5\%$.
\end{abstract}

}

\maketitle

\IEEEdisplaynontitleabstractindextext

\IEEEpeerreviewmaketitle

\justifying

\IEEEraisesectionheading{\section{Introduction}\label{sec:introduction}}

\IEEEPARstart{M}{alware} attacks against enterprises have proliferated at an alarming scale. Industry analysis reports almost 17 million malware programs targeting businesses in 2021, with estimated financial losses in billions~\cite{avtest:2021:report}. The ramifications of these attacks range from user-annoying popups to ex-filtration of sensitive data, financial loss, extortion, and even sabotaging critical infrastructures. Accordingly, malware programs can be grouped into classes based on their objectives and functionalities -- Potentially Unwanted Applications (PUA) pop up unwelcome advertisements; Bankers stealthily steal financial credentials; Backdoors open hidden access paths for a remote adversary; Spyware stealthily exfiltrates sensitive data of its victim; Downloaders install a malicious payload; Cryptominers mine cryptocurrencies for the adversary while Ransomware encrypts the data victim for extortion.

\begin{figure}[!t]
\small
    \centering
    \includegraphics[width=\linewidth]{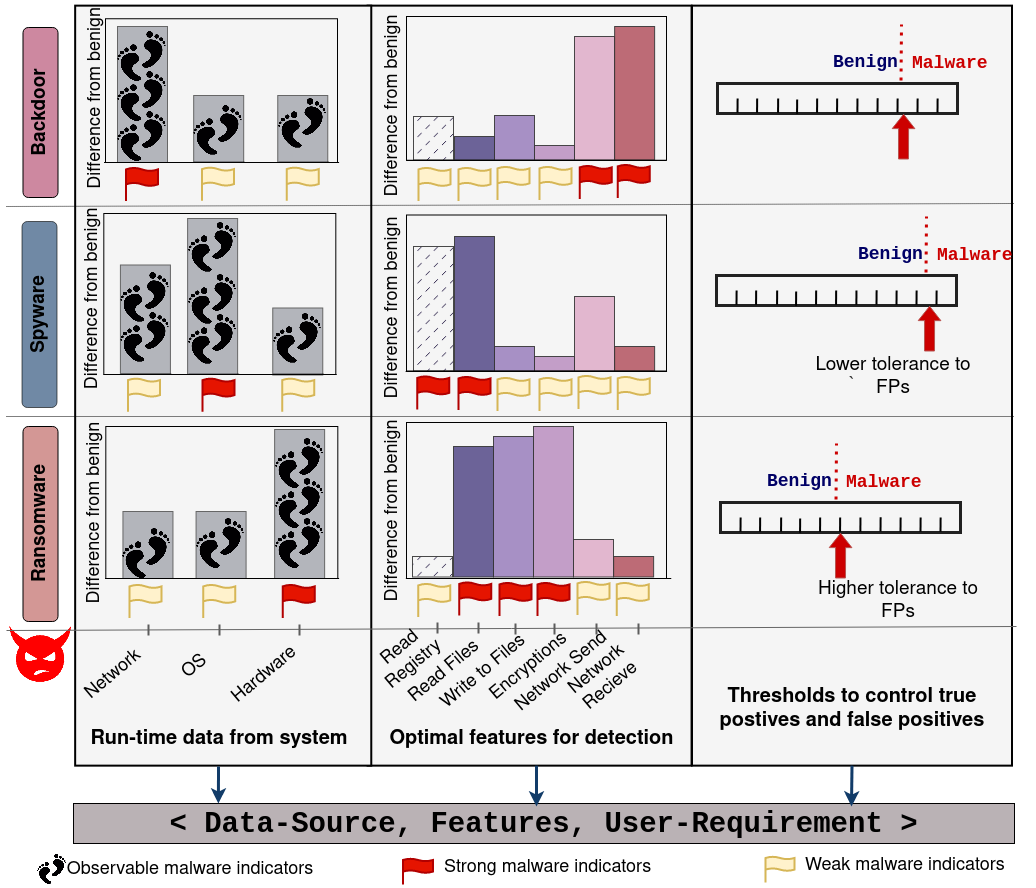}
    \vspace{-0.5cm}
    \caption{ \small The optimal tuple $\langle \tt data$-$\tt source,$ $\tt features,$ $\tt user$-$\tt requirements\rangle$ varies for each malware class. The optimal run-time $\tt data$-$\tt source$ and $\tt features$ that can distinguish a malware class from benign applications differ based on the malware functionality. Further, low-risk malware (e.g., spyware) typically have stricter classification thresholds than high-risk malware (e.g., ransomware). }
    \vspace{-0.7cm}
    \label{fig:sundewHL}
\end{figure}

The diversity in malware classes  can impact the {\em data-source}, {\em features} and the {\em user-requirements} that are instrumental in analyzing and detecting malware, as illustrated in Figure~\ref{fig:sundewHL}. First, the optimal run-time {\em data-source} that can detect a malware class differs based on the functionality. Backdoors maintain consistent communication with a remote adversary, leaving strong indicators on the network, whereas spyware are likely to leave indicators on the operating system (OS) when they scan a large number of files. On the other hand, ransomware are prone to trigger distinct hardware events due to the encryption they perform. Second, malware classes differ in the {\em features} that best identify them. For example, in the OS, a high number of encryptions and write-to files are indicators of ransomware, whereas a high rate of file-system and registry reads are indicators of spyware. Third, the {\em user's requirements} vary for different malware classes. For high-risk malware like ransomware, users are more likely to tolerate false positives than low-risk malware like spyware and PUAs.
Thus, a model for high-risk malware would ideally need lower classification thresholds than low-risk malware. Hence, detecting a malware class can benefit in accuracy and false-positive guarantees with models trained with the optimal data-source (network, OS, or hardware) and fine-tuned for class-specific features and classification thresholds. In fact, we observe that the optimal tuple of $\langle \tt data$-$\tt source, features, user$-$\tt requirements \rangle$ is unique for each malware class. Additionally, the optimal tuple is also sensitive to the system load conditions, which can infiltrate noise into the run-time data.

\begin{figure}[!t]
    \centering
    \small 
    \includegraphics[width=\linewidth]{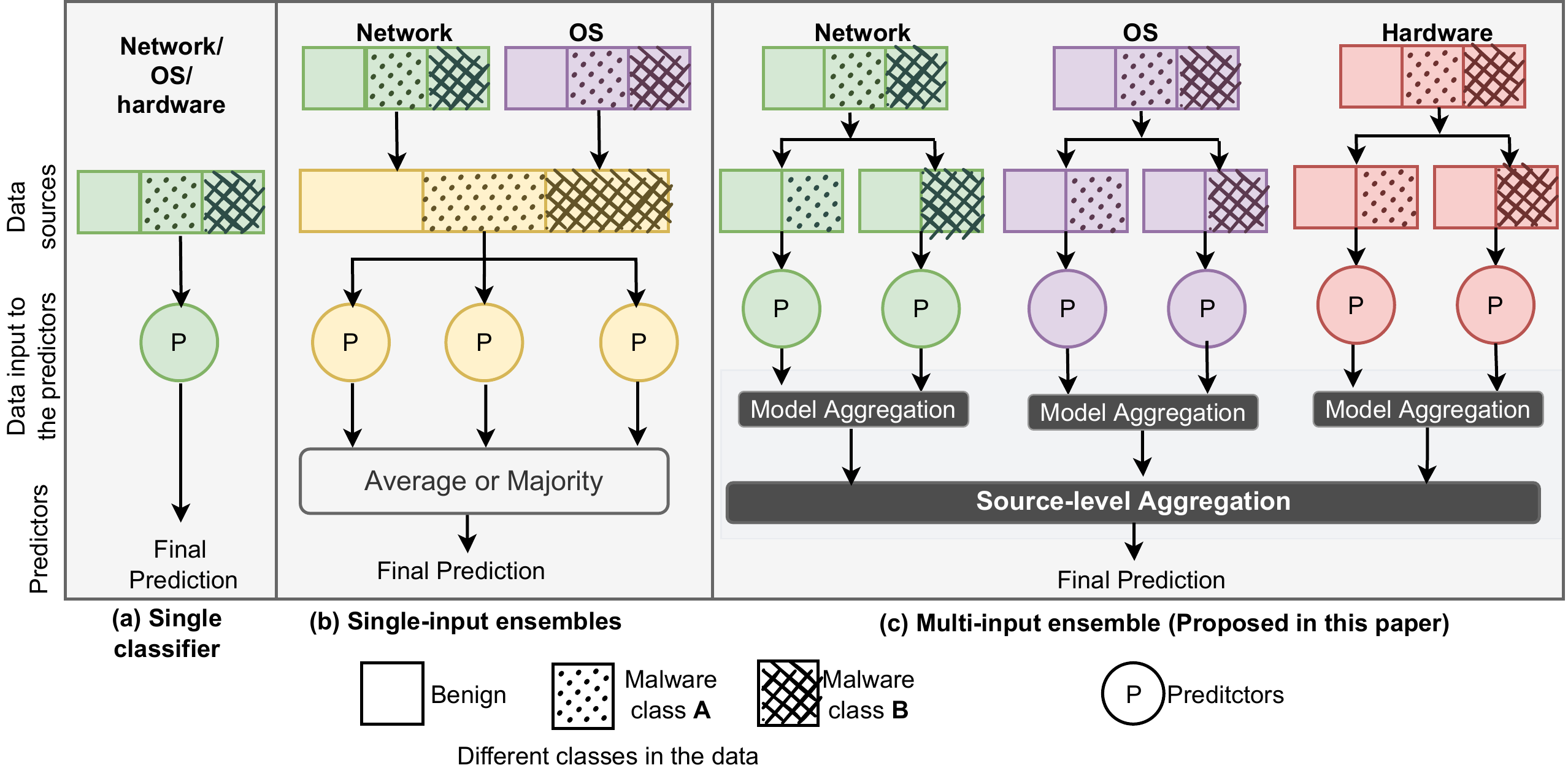}\vspace{-0.3cm}
    \caption{\small Malware detection mechanisms based on (a) single classifier; (b) single-input ensembles; and, (c) multi-input ensemble (proposed in this paper), for an example case of three classes (benign, malware classes A and B).}\vspace{-0.3cm}
    \label{fig:ensemble}
\end{figure}

The intuitive approach to leverage the optimal tuple for any malware class is to have different specialized predictors, wherein each predictor is fine-tuned for a specific data-source, class-specific features, and requirements. Na\"{i}ve solutions include hyper-specialized predictors for classes such as ransomware [32], which fail on other classes of malware. On the other hand, most works in literature employ single generic classifiers, that are too generalized to support such a specialized handling of different classes (Refer to Figure~\ref{fig:ensemble}a)~\cite{Bartos:2016:invariantrep,Wang:2013:rootkitsHPC,narudin2016evaluation, alam2013random,  anderson2017machine,nissim2014novel,Demme:2013:feasibility,Peng:2016:HPC_malware_detection_features,Tang:2014:HPCanomaly,Bayer:2009:malwareClusteringOS,watson2015malware}. Alternative works explore an ensemble-based classifiers for detection~\cite{zhou2020building, alzaylaee2020dl,guofei:2008:botminer,perdisci:2010:behavioral, feng2018novel,li2021framework,Das:2016:semantic-based_Malware_detection,sayadi2018ensemble,rhode2018early,salo2019dimensionality,tanmoy:2020:ec2}. Such solutions train a collection of predictors either in parallel or sequentially on the {\em same input data} formed by combining data from one~\cite{zhou2020building,  alzaylaee2020dl,guofei:2008:botminer,perdisci:2010:behavioral, feng2018novel,li2021framework,Das:2016:semantic-based_Malware_detection,sayadi2018ensemble} or multiple sources~\cite{rhode2018early,salo2019dimensionality,tanmoy:2020:ec2}. Figure~\ref{fig:ensemble}b illustrates one such single-input ensemble that trains predictors in parallel. Primarily, these approaches aim to minimize the detection error by averaging the predictions from the individual predictors (when trained in parallel) or learning adaptively (when trained in sequence). Either way, {\em same-input ensembles} cannot achieve the optimal tuple for detection as they do not support fine-tuning individual predictors to pursue class-specific features and requirements.

Figure~\ref{fig:ensemble}c illustrates our proposed ensemble, which we call a {\em multi-input ensemble} of specialized predictors, wherein each predictor is trained using a different input data-source to detect a specific malware class from benign applications. However, given an unknown program sample, such fine-grained specialization alone is not sufficient to leverage the optimal predictor (i.e., achieve the optimal tuple) in the ensemble. The primary challenge lies in arriving at a {\em consensus} from the independent, conflicting predictions from different specialized predictors for any input test program. As each predictor is tuned differently, their inferences are likely to conflict. Na\"{i}ve approaches such as majority or average of individual predictions~\cite{zhou2020building,sayadi2018ensemble,feng2018novel,li2021framework} may not be optimal. This is because each predictor has different definitions of boundaries between malware and benign behavior and deals with different noise levels in the input from different data sources. 

This paper presents \nameA, a detection framework that employs a {\em multi-input ensemble} of predictors with an {\em insightful aggregation} mechanism to leverage the optimal tuple of $\langle \tt data$-$\tt source, features, user$-$\tt requirements \rangle$ for any input test program. \name has three {\em components}, each utilizing different data sources across the system stack to provide a holistic view of malware activity. Internally, each component has multiple specialized predictors, each tuned to differentiate a specific malware class from benign applications using data collected from the particular data source (network, OS, or hardware) (Refer to Figure~\ref{fig:ensemble}c). To resolve the conflicts between the predictors, SUNDEW uses a two-level hierarchical structure of aggregator functions that first collates the inferences from different specialized predictors inside each component and later aggregates the inferences from the three data sources (Refer to Figure~\ref{fig:ensemble}c). These functions employ a combination of predictor statistics, prior knowledge of the capabilities of each predictor, risk factor, and
the current system load to aggregate inferences to an optimal prediction. Thus, unlike prior works, the holistic view of malware activity and specialization enable \name to achieve a case-sensitive analysis and detection, thus improving the accuracy and resilience while ensuring class-specific false-positive guarantees. 
Following are the contributions of the paper:
\vspace{-0.3cm}

\begin{enumerate}

        \item A reliable malware analysis framework, \nameA,  with a holistic view of malware activity across the system, an ensemble of specialized predictors, and aggregator functions to help derive the best-case prediction for any malware class (Section~\ref{sec:sundew}).  
    
    \item An evaluation of various design choices for the aggregator functions that resolve conflicts between the independent specialized predictors. Given an unknown sample, the two-level aggregation in \name relays the inference of the specialized predictor to the output without any loss, while boosting the performance of the optimal predictor by at least 1.42\% (Section~\ref{sec:aggregation}).

    \item An evaluation of \name on a rich dataset that presents precise and comprehensive real-world malware behavior of more than $10,000$ malware samples, including cryptominers, bankers, spyware, backdoors, ransomware, downloaders, deceptors, and potentially unwanted applications (PUAs). \name can achieve an F1-Score of $1$ for most malware classes, an average of $0.93$ for any malware class, and $0.82$ even under highly noisy conditions, with an average overhead as low as $1.5\%$ at the end-host machines (Section~\ref{sec:implementation}).

    \item To the best of our knowledge, \name is the first to provide a multi-input ensemble for a case-sensitive detection of malware classes. It evaluates the run-time inputs from three system components, risk factors, and the dynamic system noise, to detect malware reliably. \name is 10\% more accurate, with 89\% lower false positives, than prior state-of-the-art predictors based on network~\cite{Bartos:2016:invariantrep}, OS~\cite{ Das:2016:semantic-based_Malware_detection}, hardware~\cite{sayadi2018ensemble} and ensembles~\cite{tanmoy:2020:ec2} that do not consider a holistic view of malware activity and multi-input ensemble of specialized predictors \cite{guofei:2008:botminer,perdisci:2010:behavioral,Bartos:2016:invariantrep, Shanshan:2018:androidmalware,narudin2016evaluation,anderson2017machine,alzaylaee2020dl,comar2013combining,nissim2014novel,zhou2020building,sayadi2018ensemble,Bahador:2014:HPCMalHunter_SVD,Demme:2013:feasibility,Peng:2016:HPC_malware_detection_features,Tang:2014:HPCanomaly,Wang:2013:rootkitsHPC,Wang:2016:HPCMalware,feng2018novel,li2021framework,Bayer:2009:malwareClusteringOS,Canali:2012:call-based-malware-detection,Das:2016:semantic-based_Malware_detection,alam2013random,rhode2018early,salo2019dimensionality, watson2015malware,tanmoy:2020:ec2}.

\end{enumerate}

Following is the organization of the rest of the paper. Section~\ref{sec:background} provides the necessary background for the paper. Section~\ref{sec:motivation} highlights the motivation for the need for an ensemble of predictors. Section~\ref{sec:relatedwork} presents the related work. We discuss the high-level overview of~\name in Section~\ref{sec:sundew}. Section~\ref{sec:aggregation} discusses the use of different insights to effectively aggregate predictions in \nameA. Section~\ref{sec:implementation} presents the implementation of \name and results of our evaluation. Section~\ref{sec:discussion} discusses the limitations of~\name and future work. Finally, Section~\ref{sec:conclusion} concludes the paper.

\section{Background}
\label{sec:background}

Malware are programs with malicious intents. With differing attack objectives, they pose varying levels of risk to system users. Ransomware that can sabotage an entire system is a high-risk malware, whereas adware or potentially unwanted applications (PUA) that are mere user-annoying in nature are low-risk malware~\cite{risklevel}. Table~\ref{tab:malware_objectives} presents a few notable malware classes with their objectives and corresponding risk levels.

 {\flushleft \bf Run-time data sources for malware detection.} Malware behavioral analysis and detection is a widely studied and mature field~\cite{alam2019ratafia, guofei:2008:botminer,perdisci:2010:behavioral,Bartos:2016:invariantrep, Shanshan:2018:androidmalware,narudin2016evaluation,anderson2017machine,alzaylaee2020dl,comar2013combining,nissim2014novel,zhou2020building,sayadi2018ensemble,Bahador:2014:HPCMalHunter_SVD,Demme:2013:feasibility,Peng:2016:HPC_malware_detection_features,Tang:2014:HPCanomaly,Wang:2013:rootkitsHPC,Wang:2016:HPCMalware,feng2018novel,li2021framework,Bayer:2009:malwareClusteringOS,Canali:2012:call-based-malware-detection,Das:2016:semantic-based_Malware_detection,alam2013random,rhode2018early,salo2019dimensionality, watson2015malware,tanmoy:2020:ec2}.
The detection mechanisms use trails of malware activity, observable at different system components which include -- \textbf{(1)} Network (e.g. malware communications to its command-and-control server); \textbf{(2)} Operating system  (e.g. system calls); and, \textbf{(3)} Hardware (e.g. micro-architectural events). 

These behavioral trails provide different abstractions of malware behavior. The network trails provide insights into malware communications to external entities, including its command-and-control servers. The features of interest include the unencrypted meta-data about connections, domains contacted, TLS handshakes, and X.509 certificates from HTTPS/HTTP flows. In contrast, the OS component captures the malware interactions with the system software when it attempts to remain stealthy, achieve persistence, and execute its objective. These interactions include the file system, registry, process, and network-related system call traces of the malware. Though the system call traces include network communications (TCP Send/Receive), they are at a higher abstraction as compared to that captured at the network component. Additionally, malware activities also trigger specific micro-architectural events that can be observed using special registers called Hardware Performance Counters (HPCs)~\cite{intel:2008:intelMnaual3B}. A few notable HPC events used for malware detection include cache hits/misses at various levels, branch instruction response, and CPU activity. These behavioral trails are used to train detection models that predominately rely on machine learning (ML) to differentiate malicious and benign behavior.

\begin{table}[t!]
    \centering
    \small 
    \caption{ \small  Objectives and risk levels of various malware classes}
    \scriptsize
    \begin{tabular}{|M{0.15\linewidth}|M{0.4\linewidth}|M{0.09\linewidth}|M{0.13\linewidth}|}
    
    \hline
    {\bf Malware Class} & {\bf Objectives} & {\bf Risk level~\cite{risklevel}} & {\bf User requirements }\\
    \hline
    Cryptominer & Exploit computing resources of the victim to mine cryptocurrencies & High & High TPR\\
    \hline
    Banker & Steal financial credentials & High & High TPR \\
    \hline
    Spyware & Infiltrate and keep gleaning sensitive information for an extended period & Medium & Low FPR\\
    \hline
    Backdoor & Grant alternate covert pathways to system resources, bypassing access control & Very High & High TPR\\
    \hline
    Ransomware & Sabotage user files and extort a ransom from the user for restoration & Very High & High TPR\\
    \hline
    PUA & Pop-up annoying advertisements and inappropriate content & Low & Low FPR\\
    \hline
    Downloader & Covertly download other malware from a remote server to execute and infect & Low & Low FPR\\
    \hline
    Deceptor & Bypass the system with close to benign behavior and adware-like payload & Low & Low FPR\\
    \hline
    \end{tabular}
    \label{tab:malware_objectives}
\end{table}

{\flushleft \bf Relevance of HPC in malware detection.} 
Though the use of HPCs for malware detection is much debated, they are found to be beneficial when used in the right way, using interrupt and context switch management~\cite{das2019sok}. Additionally, with minimal instrumentation required in the HPC trails before they can be fed to the models and their limited performance overheads, HPC-based detection enables a trade-off between accuracy and overhead in malware detection.

{\flushleft \bf User Requirements.} The detection models are fine-tuned for an optimal trade-off between two orthogonal user requirements -- \textbf{(1)} a high true-positive rate (TPR), with some tolerance to mispredictions, to detect as many malware as possible; or a \textbf{(2)} a low false-positive rate (FPR), with no tolerance to mispredictions, to prevent any impact on benign applications. These requirements are, in turn, dependent on the risk level of the malware. Users would prioritize a high TPR for high-risk malware while preferring a low FPR for adware/PUA that are very similar to benign applications in behavior. Table~\ref{tab:malware_objectives} provides preferred user requirements based on the risk level of different malware classes.

\section{Motivation}
\label{sec:motivation}

In this section, we analyze the differences in malware run-time activity and present our observations that motivate the need for a case-sensitive analysis of malware. To this end, we study the difference in the behavior of malware classes from benign applications observable across the three data-sources. For each class, we assess the difference using specialized binary ML models, each trained with behavioral data of $1000$ programs, including the malware class and benign applications. Based on our observations, we make the following claims on the benefits of exploring multiple data-sources and class-specific features and addressing user-specific requirements.

\begin{figure}[t]
    \centering
    \small 
  \includegraphics[scale=0.32]{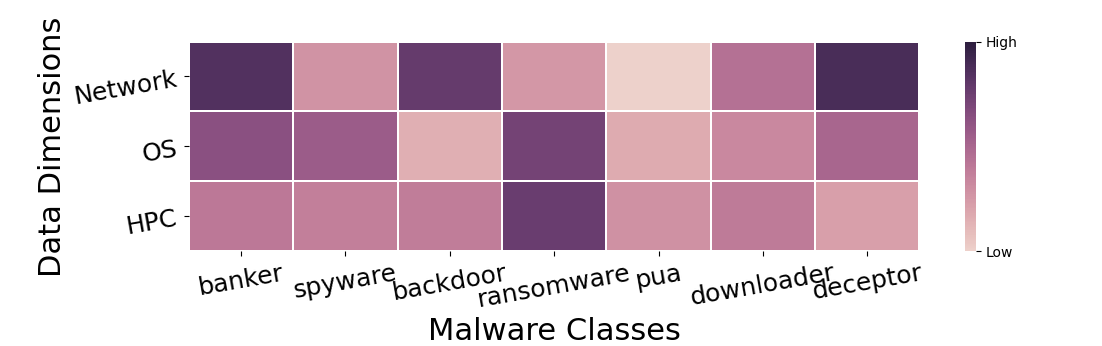} 
\vspace{-5mm}
  \caption{ \footnotesize  Differences between malware and benign behavior for different classes across the three data data-sources. The darker the color, the more distinguishable are the activities from benign applications.}
    \label{fig:motivation}
    \vspace{-4mm}

\end{figure}

{\flushleft \bf C-1: Detection can benefit in performance from a cross-dimensional view of malware activity.} Malware classes differ in their actions. In turn, the actions determine the quantum of malware activity across the system:  network, OS, and hardware. Figure~\ref{fig:motivation} indicates the distinguishability of activities of different malware classes from benign applications across the three data-sources. The darker the color, the more distinguishable are the activities from benign applications. The network shows a high distinction in activities for banker, backdoor, and deceptor, whereas OS shows distinguishable activity for ransomware and spyware. Similarly, hardware shows distinct activity in the case of ransomware. Hence, some classes are better aligned to be detected using a specific data-source than others. We explain this with an example of backdoor, spyware, and ransomware (Refer to Figure~\ref{fig:sundewHL}). A backdoor creates a reverse shell, escalates privileges, and provides code injection capabilities to a remote adversary. However, the constant factor in its attempt to execute any command from the adversary is its sustained communication with the remote server. We observe that the average duration of network flows for a backdoor is notably different from other classes. On the other hand, spyware aims to gather information about the victim. Hence, it scans the filesystem, leaving distinct trails at the OS, while its network activities are not significantly distinguishable from benign applications. Similarly, ransomware scans the files at the victim to encrypt and make them inaccessible. In the process, a high rate of reading and writing files is observable at the OS. However, the high encryption rate leaves a significant fingerprint of micro-architectural events visible at the hardware. Hence, it is beneficial to {\em explore multiple data-sources to get a complete picture of malware activity}, and build appropriate defenses.  

{\flushleft \bf C-2: Detection can benefit in resilience by employing all data-sources.}
The data collected at the network and hardware is affected by other processes executing in the system. With an increase in system load (i.e., the number of processes), the network communications of other processes get induced into the network data. Similarly, other processes sharing micro-architectural resources in the system can affect the hardware performance counters. In contrast, the OS data is collected only for the specific PID and hence is agnostic to the system load. Thus, it is beneficial to {\em employ multiple data-sources for resilient malware detection. }

{\flushleft \bf C-3: Detection can benefit from class-specific features.} Within each data-source, malware classes differ in the features that best express their maliciousness. As an example, in hardware, the performance counter event, which counts the number of switches from the Decode Stream Buffer (DSB) to the Micro-instruction Translation Engine (MITE), $\tt DSB2MIT\_SW\_CNT$ is empirically one of the most important features in classifying ransomware from benign applications. However, for spyware, the corresponding feature is the event $\tt M\_LD\_Ret\_L1Hit$, which counts retired loads that encounter a hit in the L1 cache in a specific cache coherency state. Thus, {\em it is beneficial to have predictors specialized for class-specific features rather than a generic approach.}

{\flushleft \bf C-4: Detection can benefit if fine-tuned to appropriate user requirements.} System users respond differently to different malware classes. For instance, users would want to kill ransomware as soon as possible to restrict further damage. Hence, models for high-risk malware (e.g., ransomware, backdoor) target a high true-positive rate to detect every malware sample, while tolerating some false positives. In contrast, users are comparatively lenient to low-risk malware (e.g., PUA and deceptor) that are mere user annoying in nature. Users would prefer to kill such malware only if the prediction is precise to reduce any impact on benign applications. Thus, false positives are a concern for such malware. The predictor classification threshold controls the trade-off between the true-positive rate (TPR) and the false-positive rate (FPR). 
High-risk malware would require lower thresholds to promote high TPR, whereas low-risk malware would require stricter thresholds to reduce FPR. Hence, {\em it is beneficial to have predictors specialized for class-specific user requirements.}

In essence, the optimal tuple $\langle \tt data$-$\tt source, features, user$-$\tt requirement \rangle$ for each malware class is different, making it essential to have a holistic view of the data-sources and specialized models to improve detection efficiency.

\section{Related Work}
\label{sec:relatedwork}
\begin{figure}
    \centering
    \small 
    \includegraphics[width=\columnwidth]{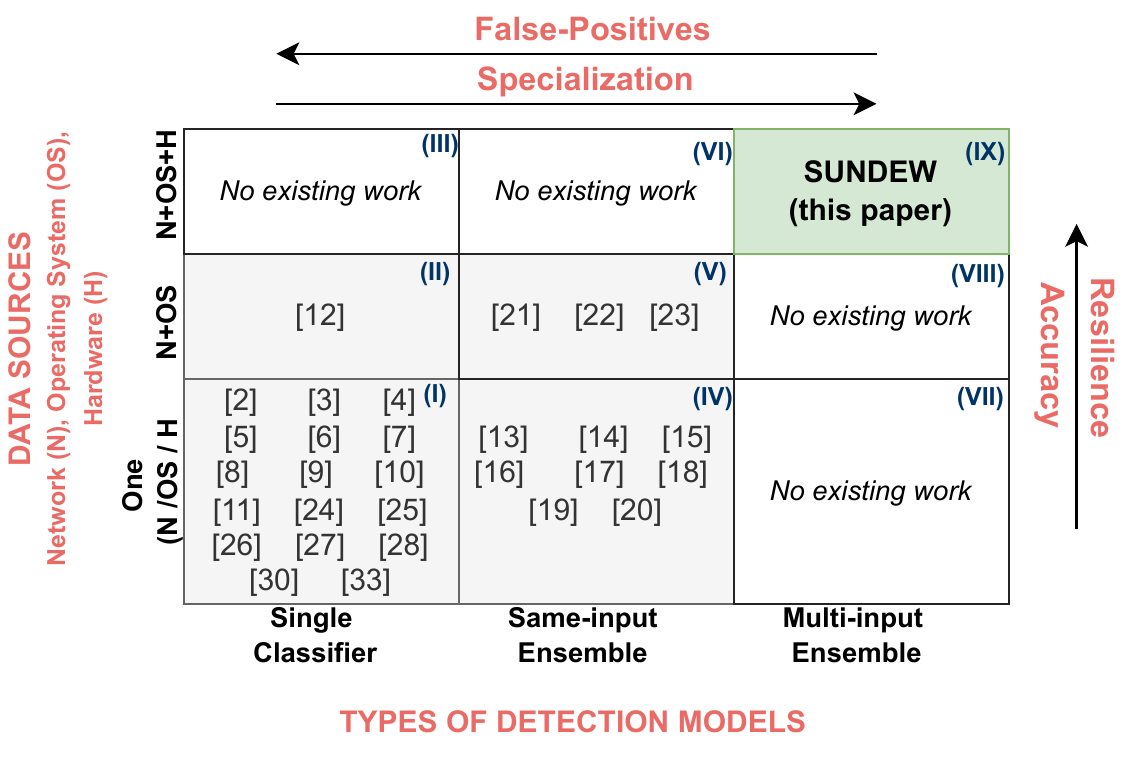}
        \vspace{-9mm}

\caption{\footnotesize Prior detection mechanisms differ in specialized handling of malware classes based on the run-time data source and the machine learning models they employ.}
\label{tab:relwork}
    \vspace{-3mm}

\end{figure}

Malware analysis and detection using run-time behavior have been extensively explored in literature~\cite{alam2019ratafia, guofei:2008:botminer,perdisci:2010:behavioral,Bartos:2016:invariantrep, Shanshan:2018:androidmalware,narudin2016evaluation,anderson2017machine,alzaylaee2020dl,comar2013combining,nissim2014novel,zhou2020building,sayadi2018ensemble,Bahador:2014:HPCMalHunter_SVD,Demme:2013:feasibility,Peng:2016:HPC_malware_detection_features,Tang:2014:HPCanomaly,Wang:2013:rootkitsHPC,Wang:2016:HPCMalware,feng2018novel,li2021framework,Bayer:2009:malwareClusteringOS,Canali:2012:call-based-malware-detection,Das:2016:semantic-based_Malware_detection,alam2013random,rhode2018early,salo2019dimensionality, watson2015malware,tanmoy:2020:ec2, Khasawneh:2017:RHMD}. Figure~\ref{tab:relwork} compares the highly cited prior works in the last decade based on the run-time data sources and the machine learning models they employ to cater to different malware classes.

{\flushleft \bf Run-time data sources.} Most prior works employ run-time data from a {\em single} system component to detect malware (refer to cells I and IV in Figure~\ref{tab:relwork}). These include trails from the network~\cite{guofei:2008:botminer,perdisci:2010:behavioral,Bartos:2016:invariantrep, Shanshan:2018:androidmalware,narudin2016evaluation, anderson2017machine,alzaylaee2020dl,comar2013combining,nissim2014novel,zhou2020building}, operating system (OS)~\cite{Bayer:2009:malwareClusteringOS,Canali:2012:call-based-malware-detection,Das:2016:semantic-based_Malware_detection,alam2013random}, or hardware~\cite{alam2019ratafia, sayadi2018ensemble, Bahador:2014:HPCMalHunter_SVD,Demme:2013:feasibility,Peng:2016:HPC_malware_detection_features,Tang:2014:HPCanomaly,Wang:2013:rootkitsHPC,Wang:2016:HPCMalware,feng2018novel,li2021framework, Khasawneh:2017:RHMD}. Alternatively, few works employ a combination of features extracted from the network and OS (refer to cells II and V in Figure~\ref{tab:relwork}~\cite{rhode2018early,salo2019dimensionality, watson2015malware,tanmoy:2020:ec2}). While the multiple inputs from network and OS can increase the detection accuracy, these solutions can miss indicators of classes like ransomware that have high micro-architectural activity (refer to C-1 in Section~\ref{sec:motivation}). Unlike prior solutions, SUNDEW employs a comprehensive view of malware activity across the computing stack (cell IX), including hardware, thus improving the accuracy and resilience of detection (refer to C-1 and C-2 in Section~\ref{sec:motivation}).

{\flushleft \bf Generic vs. Specialized models.} State-of-the-art solutions in dynamic analysis use different detection models, as shown in Figure~\ref{fig:ensemble} and in the columns of Figure~\ref{tab:relwork}. While these models can provide a binary or multi-class prediction, internally, they either use a {\em single} classifier (cells I and II) or a {\em same-input ensemble} of predictors (cells IV and V). A single classifier can use data from a single (cell I~\cite{narudin2016evaluation, anderson2017machine,nissim2014novel,Bartos:2016:invariantrep,Shanshan:2018:androidmalware,comar2013combining,alam2013random, Bayer:2009:malwareClusteringOS, Khasawneh:2017:RHMD,  Bahador:2014:HPCMalHunter_SVD,Peng:2016:HPC_malware_detection_features, Demme:2013:feasibility, Tang:2014:HPCanomaly, Wang:2016:HPCMalware,Wang:2013:rootkitsHPC, Canali:2012:call-based-malware-detection, alam2019ratafia}) or multiple sources (cell II~\cite{watson2015malware}). Though relatively light-weight, most single predictors are too generalized to support any specialization~\cite{narudin2016evaluation, anderson2017machine,nissim2014novel,Bartos:2016:invariantrep,Shanshan:2018:androidmalware,comar2013combining,alam2013random, Bayer:2009:malwareClusteringOS, Khasawneh:2017:RHMD,  Bahador:2014:HPCMalHunter_SVD,Peng:2016:HPC_malware_detection_features, Demme:2013:feasibility, Tang:2014:HPCanomaly, Wang:2016:HPCMalware,Wang:2013:rootkitsHPC, Canali:2012:call-based-malware-detection, watson2015malware}. On the other hand,  hyper-specialized predictors~\cite{alam2019ratafia} that are fine-tuned for specific classes, such as ransomware, can fail on other classes of malware.

Alternatively, a {\em same-input} ensemble (cell IV) consists of multiple predictors, wherein all predictors train on the same data. These predictors are trained either in sequence or parallel to improve the predictive performance~\cite{zhou2020building,  alzaylaee2020dl,guofei:2008:botminer,perdisci:2010:behavioral, feng2018novel,li2021framework,Das:2016:semantic-based_Malware_detection,sayadi2018ensemble}. While the former approach trains the predictors sequentially in an adaptive manner~\cite{alzaylaee2020dl,guofei:2008:botminer,perdisci:2010:behavioral,Das:2016:semantic-based_Malware_detection}, the latter employs all predictors in parallel and outputs the average of their predictions~\cite{zhou2020building,sayadi2018ensemble,feng2018novel,li2021framework}. Inherently, such ensembles do not support training predictors with different data pertaining to a specific class and benign behavior. The addition of alternate data sources \cite{rhode2018early,salo2019dimensionality,tanmoy:2020:ec2} (cell V) does not help either, as the features from different input sources are transformed to a single representation before feeding to the model. Unlike prior works, \name proposes a {\em multi-input} ensemble of predictors, wherein each classifier trains on a different run-time data source and malware class and is specialized to maximize the class-specific user requirement (cell IX).

 When predictors in a same-input ensemble (that are trained in parallel) test any given unknown sample, their independent predictions are {\em aggregated} (e.g., by averaging) to minimize the cumulative errors of the individual predictors and form a final prediction~\cite{zhou2020building,sayadi2018ensemble,feng2018novel,li2021framework, salo2019dimensionality}. In contrast, aggregation in the multi-input ensemble of \nameA, aims to relay the inference of the specialized predictor corresponding to the given sample to the final output. Challenges in such an aggregation are two-fold. First, the individual predictors in a multi-input ensemble deal with different noise levels in the input data from different data sources. Second, the predictors have different definitions of positive and negative class. We discuss how \name addresses these aggregation challenges to relay the optimal prediction of the specialized predictor to the final output in Section~\ref{sec:aggregation}. Other industrial solutions such as~\cite{joesandbox} present class-specific behavioral analysis similar to the goals of \nameA. However, based on our limited understanding, these closed-box solutions do not support aggregation to present a final prediction of the input sample.

Thus, the comprehensive view of malware run-time activity, the multi-input ensemble and the aggregation mechanism ensure that \name is \textbf{(1)} more accurate in detecting malware (unlike~\cite{guofei:2008:botminer,perdisci:2010:behavioral,Bartos:2016:invariantrep, Shanshan:2018:androidmalware,narudin2016evaluation,anderson2017machine,alzaylaee2020dl,comar2013combining,nissim2014novel,zhou2020building,sayadi2018ensemble,Bahador:2014:HPCMalHunter_SVD,Demme:2013:feasibility,Peng:2016:HPC_malware_detection_features,Tang:2014:HPCanomaly,Wang:2013:rootkitsHPC,Wang:2016:HPCMalware,feng2018novel,li2021framework,Bayer:2009:malwareClusteringOS,Canali:2012:call-based-malware-detection,Das:2016:semantic-based_Malware_detection,alam2013random,rhode2018early,salo2019dimensionality, watson2015malware,tanmoy:2020:ec2, Khasawneh:2017:RHMD} that do not exploit C-1 and C-3); \textbf{(2)} more resilient to infiltrating noise (unlike~\cite{guofei:2008:botminer,perdisci:2010:behavioral,Bartos:2016:invariantrep, Shanshan:2018:androidmalware,narudin2016evaluation,anderson2017machine,alzaylaee2020dl,comar2013combining,nissim2014novel,zhou2020building,sayadi2018ensemble,Bahador:2014:HPCMalHunter_SVD,Demme:2013:feasibility,Peng:2016:HPC_malware_detection_features,Tang:2014:HPCanomaly,Wang:2013:rootkitsHPC,Wang:2016:HPCMalware,feng2018novel,li2021framework} that do not exploit C-2); \textbf{(3)} can support class-specific false positives (unlike~\cite{guofei:2008:botminer,perdisci:2010:behavioral,Bartos:2016:invariantrep, Shanshan:2018:androidmalware,narudin2016evaluation,anderson2017machine,alzaylaee2020dl,comar2013combining,nissim2014novel,zhou2020building,sayadi2018ensemble,Bahador:2014:HPCMalHunter_SVD,Demme:2013:feasibility,Peng:2016:HPC_malware_detection_features,Tang:2014:HPCanomaly,Wang:2013:rootkitsHPC,Wang:2016:HPCMalware,feng2018novel,li2021framework,Bayer:2009:malwareClusteringOS,Canali:2012:call-based-malware-detection,Das:2016:semantic-based_Malware_detection,alam2013random,rhode2018early,salo2019dimensionality, watson2015malware,tanmoy:2020:ec2} that do not exploit C-4); and, \textbf{(4)} caters to a diverse set of malware classes (unlike~\cite{alam2019ratafia}). 

{\flushleft \bf Comparison with SIEM.} The comprehensive analysis in \name has similarities with industrial efforts such as System Information and Event Management (SIEM)~\cite{miller2011security} that consider a holistic correlation of events across an enterprise to detect threat scenarios. However, there are differences between the two. First, SIEM collects data from diverse sources such as antivirus software, security appliances, firewalls, and other organizational applications, including human interactions (such as repeated access attempt failures). It correlates these events against pre-defined rules to detect threats and create alerts. In contrast, \name is an alternative input source to SIEM that can replace the antivirus software to provide accurate and resilient detection of malware activities. Second, the machine learning ensemble in \name is more sophisticated than the correlation rules predominantly used in SIEM and thus capable of detecting zero-day threat scenarios.

{\flushleft \bf Comparison with multi-input solutions in static analysis.} Few prior efforts based on static analysis have explored malware detection using multi-input ensembles\cite{kim2018multimodal, yan2018detecting}. These solutions employ a heterogeneous ensemble of predictors, where each classifier trains on different static features extracted from the malware binary. However, unlike dynamic analysis, static techniques can be easily evaded by polymorphic and metamorphic malware that are popular today. To the best of our knowledge,~\name is the first multi-input ensemble for dynamic analysis, considering a comprehensive view of run-time activity to provide a case-sensitive detection of malware.

\section{The \name Framework}
\label{sec:sundew}

In this section, we first present a high-level overview of~\name followed by a formal description of the multi-input ensemble.

\subsection{High-Level Overview}
\begin{figure}[t]
    \centering
    \small
    \includegraphics[scale=0.2]{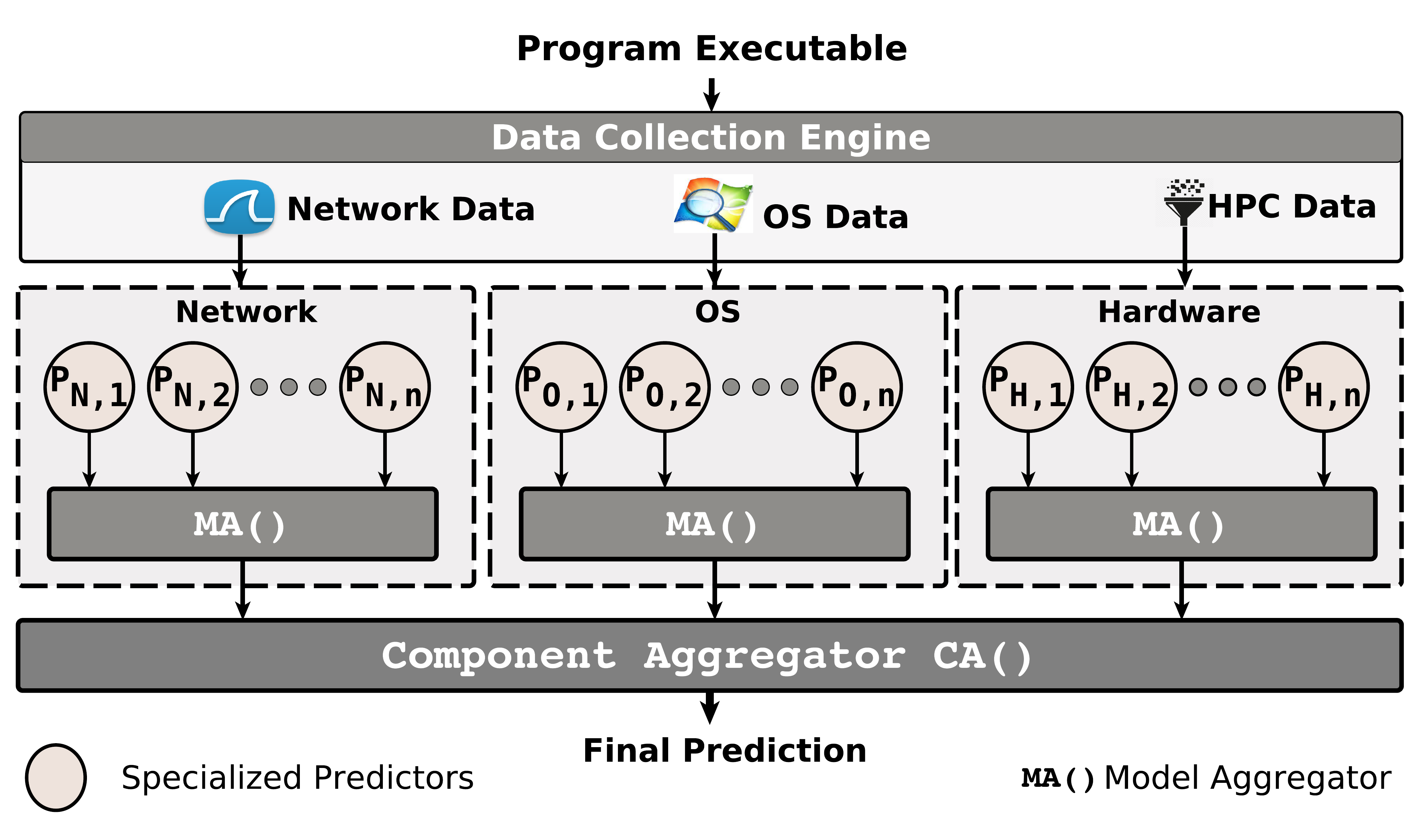}
        \vspace{-3mm}

    \caption{\small \textbf{Multi-input ensemble of predictors in \nameA}. \footnotesize The ensemble consists of three components corresponding to each data source (network, OS, hardware). Internally, each component has a specialized predictor for each class. $\tt MA ()$ aggregates the inferences of the specialized predictors inside each component, whereas the $\tt CA ()$ aggregates the inference of the three components to make the final output of \name.}
    \label{fig:hlv}
    \vspace{-6mm}

\end{figure}

\name is a multi-input ensemble of predictors that leverages the optimal tuple of $\langle \tt data$-$\tt source$, $\tt features$, $\tt user$-$\tt requirements \rangle$ for accurate and resilient detection of any malware class. Figure~\ref{fig:hlv} presents a high-level overview of the working of~\nameA. The ensemble has a component for each data-source, namely, network, OS, and hardware (shown by the dashed boxes in Figure~\ref{fig:hlv}). During program execution, a collection engine collects the program's network, OS, and hardware behavioral data and invokes the respective components. Internally, each component has multiple predictors, each of which is specialized for different malware classes with class-specific features and user requirements (refer to {\bf C-3} and {\bf C-4} in Section~\ref{sec:motivation}). Each predictor is a binary classifier trained for a specific class, where it can infer if the program is of the malware class of its specialization or benign. For example, a backdoor-specialized predictor predicts if a program is a backdoor, likewise, a PUA-specialized program predicts if a program is a PUA.

\begin{figure}[!t]
    \centering
    \small
    {\includegraphics[width=\linewidth]{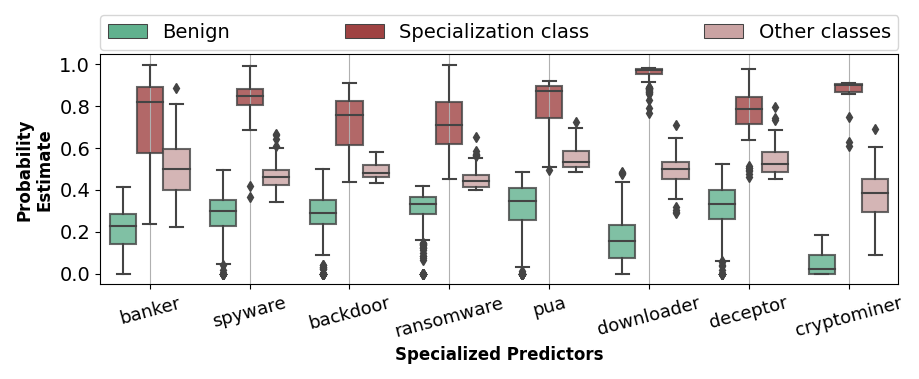} }
        \vspace{-7mm}

    \caption{\footnotesize Distribution of probability estimates of different specialized predictors in the network component, when tested with data of benign class (green), the corresponding specialization (red), or from any other class (light red). Each predictor has different definitions of class boundaries.}
        \label{fig:prob}
            \vspace{-4mm}

\end{figure}

The predictors are likely to output conflicting inferences. Primarily, each specialized predictor has a different definition of the boundary between benign and malware. Figure~\ref{fig:prob} shows the distribution of probability estimates of different specialized predictors in the network component when they are tested with benign applications (green boxes), and the class of their specialization (dark-red boxes). As evident these distributions overlap. Thus, due to similarities between deceptor and some benign applications, the backdoor-predictor might infer a deceptor program as benign while, the deceptor-predictor infers it as a deceptor. Further, while a predictor can predict its class of specialization with high confidence (dark-red boxes), it predicts other classes as malware with varying likelihood (light-red boxes). As~\name starts with no notion of the program's class; the challenge lies in choosing the right prediction from the set of independent predictions. 
\name addresses this challenge by aggregating predictions using a configurable model-aggregator within each component. The model-aggregator $\tt MA()$ multiplexes the output of the best-case specialized predictor to the output. For this, it leverages predictor statistics (e.g. probability estimates) and prior knowledge to assess the {\em confidence} of inferences inside each component. 
Prior knowledge can include the capabilities of components to reveal certain classes as observed during the training phase. For instance, prior knowledge that hardware trails have strong indicators of ransomware can help assess the confidence of a ransomware-specialized predictor in hardware. Based on the statistics and prior knowledge, $\tt MA()$ computes the confidence score of each predictor and relays the most confident inference as the output of the component.

The outputs from the three $\tt MA()$s are likely to differ on the class of the program due to two reasons. First, each data-source varies on its capability to distinguish a specific malware class from benign (Refer to Figure~\ref{fig:motivation} and Claim C-1 in Section~\ref{sec:motivation}). Second, the data-sources have varying levels of noise from other processes running in the system (Refer to Claim C-2 in Section~\ref{sec:motivation}). For this, \name uses a component-aggregator function $\tt CA()$ to aggregate the outputs from the model-aggregators and yield the most confident inference as the final classification output. Similar to $\tt MA()$, it exploits the components' statistics (the output confidence score from $\tt MA()$) and its prior knowledge. Its prior knowledge can include a broader understanding of the distinguishing capabilities of different data sources. Further, the $\tt CA()$ also checks the system load and considers the output of a resilient data source, with the least infiltrating noise for aggregation to the output. At both model and component aggregators, multiple predictors/components likely can end up with similar confidence scores. In such scenarios, both $\tt MA()$ and $\tt CA()$ leverage the known risk-level~\cite{risklevel} of the classes as a tie-breaker. They choose the riskiest class as their aggregated inference to resolve the tie. For instance, if two conflicting inferences, backdoor and deceptor have similar confidence scores, the aggregators choose the higher-risk class (backdoor) of the two.  
 
 Thus, given any input test program, relaying the predictive benefits of the corresponding specialized predictor (that employs the optimal data-source, features, and user-requirements) to the output of \name is important to improve accuracy and resilience. The choice and weights of statistics and prior knowledge control this relay and the effectiveness of the aggregation. We explore this aspect and evaluate different designs for $\tt MA()$ and $\tt CA()$ in Section~\ref{sec:aggregation}.

\subsection{Formal Description of SUNDEW}

Let $\tt \mathbb B$ = $\tt \langle \mathcal D, \mathcal M, \mathcal P , \mathcal A\rangle$  represent the \name ensemble (Refer to Figure~\ref{fig:hlv}). $ \tt \mathcal D = \{N, O, H \}$ is the set of components (i.e. data sources), namely network ($\tt N$), OS ($\tt O$) and hardware ($\tt H$). ${\mathcal M = \{\tt m_1,m_2,\dots m_n\}}$ is the set of $\tt n$ malware classes. $\mathcal P$ is the set of specialized predictors, while $\mathcal A$  is the set of aggregator functions. 
Algorithm~\ref{alg:sundew} describes \nameA. It takes as input a program $\tt z$ from the set of programs $\tt Z$ to test. For each component $\tt k$, it first gets the behavioral data for the program (Line 5).

\begin{figure}[t]
    \centering
    \small
    {\includegraphics[width=\linewidth]{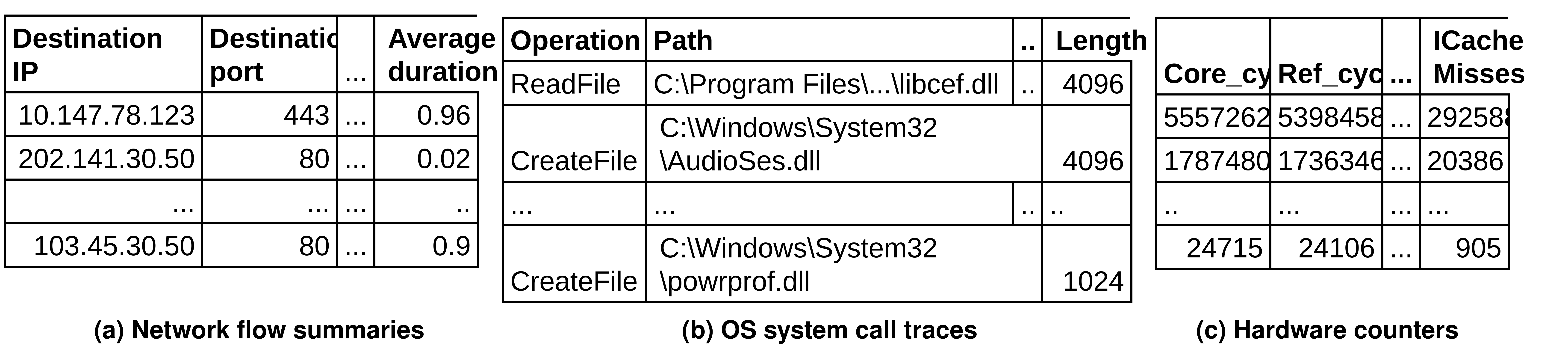} }
            \vspace{-4mm}
    \caption{\small Behavioral snapshots collected at different granularities at network, OS, and hardware components.}
        \label{fig:datasnapshot}
        \vspace{-3mm}
\end{figure}

{\flushleft \bf Behavioral data.} Given a program $\tt z$, its behavioral data in component $\tt k\in$ $\tt \mathcal D$ corresponds to a time series of snapshots collected during the execution of the program. These snapshots are captured at different granularities across components, as shown in Figure~\ref{fig:datasnapshot}. At the network, we log the data for every flow\footnote{All communications having the same source and destination IP address, and source and destination port belong to a flow. Thus the network packets are grouped into traffic flow summaries} while we log every system call at the OS. The hardware component logs the HPCs at a fixed time interval of 100ms. The collected data is pre-processed and converted to a matrix $\tt d_{z,k}$, with features as columns and rows representing each snapshot (Line 5). These rows are labeled with the class of the program, $\tt m_j$ $\tt \in $ $\tt \mathcal M, \forall j \in [1,n]$.

\begin{figure}[!t]
    \centering
    {\includegraphics[width=0.98\linewidth]{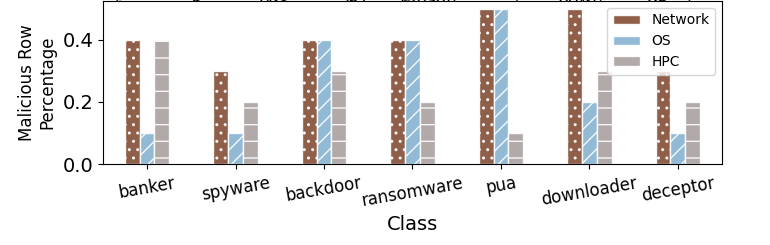} }
                \vspace{-4mm}
    \caption{\small   Percentage of malware rows per program varies for different classes, across the three components (data sources).}
                \vspace{-4mm}
        \label{fig:rows}
\end{figure}

{\flushleft \bf Prediction.} After getting the behavioral data, Algorithm~\ref{alg:sundew} invokes all the predictors in $\mathcal P$, each of which is specialized to detect one of the classes in $\mathcal M$ (Line 6). $\mathcal P$ is given by,

\begin{equation}
   \mathcal P = \{ \tt P_{k,j}(), \forall k \in \mathcal D, \forall j \in [1,n]\} \enspace,
\end{equation}

\noindent where predictor $\tt P_{k,j}$ is specialized in component $\tt k$ to classify a program as malware $\tt m_j$ or benign. These predictors are trained to predict row-wise inferences along with the probability estimate of a row in $\tt d_{z,k}$ being malicious. Accordingly, the $\tt probability$-$\tt estimates$ for $\tt d_{z,k}$ is the cumulative average of probability estimates of its rows. As the predictors are trained row-wise, the prediction of a test program might contain some rows inferred as malware (malicious) while others as benign. Interestingly, the percentage of malicious rows per program, i.e. $ \tt malicious$-$\tt row$-$\tt percentage$ varies for different classes (Refer to Figure~\ref{fig:rows}). Accordingly, the specialized predictors $\tt P_{k,j} ()$ are fine-tuned to these class-specific thresholds (in Figure~\ref{fig:rows}) to conclude the class of the test program. For instance, the network-based backdoor-specialized predictor infers a program as malware if at least 40\% of the rows are identified as malicious. Similarly, spyware-specialized predictor infers malware if at least 30\% of the rows in the program are malicious. 

Hence, for data $\tt d_{z,k}$ of a program $\tt z$ in component $\tt k$, $\tt P_{k,j}$ outputs a tuple of its prediction $\tt r_{k,j}$ and statistics $\tt s_{k,j}$ as follows:

\begin{equation}
\langle \tt r_{k,j}, s_{k,j} \rangle = 
  \tt P_{k,j}(d_{z,k}), \forall { \tt k} \in \mathcal D, \forall j \in [1,n] \enspace ,  
    \label{eqn:pkj}
\end{equation}

\noindent where,

\begin{equation}
  \tt r_{k,j} =
    \begin{cases}
      1 & \text{, for a malware of class $\tt m_j$}\\
      0 & \text{, for a benign program}\\
    \end{cases}  \enspace ,
    \label{eqn:rkj}
\end{equation}

\noindent and, $\tt s_{k,j}$ is a tuple of $\langle$ $\tt probability$-$\tt estimates$, $\tt malicious$-$\tt row$-$\tt percentage$$\rangle$. The statistics $\tt s_{k,j}$ is an indicator of {\em confidence} of the prediction of $\tt r_{k,j}$. Thus, the output from the predictors in component $\tt k$ are the set of predictions ($\tt R_k = \{r_{k,j}, \forall j \in [1,n] \}$) and statistics ($\tt S_k = \{ s_{k,j}, \forall \tt j \in \tt [1,n] \}$) (Line 7 in Algorithm~\ref{alg:sundew}). Each element in these sets corresponds to a malware class.

\begin{algorithm}[t]
\DontPrintSemicolon
\small
\KwIn{$\tt z$: Program to test 
}

\KwResult{$\tt \langle \hat{r}_{\mathbb B} , \hat{c}_{\mathbb B} \rangle$: Final label and confidence.}

\Begin{

$\tt \mathcal{D} \gets$ $\tt \{N, O, H \}$ components

$\tt Prior\-Knowledge \gets$ Prior knowledge on predictors in $\tt \mathcal P$

\tcc {\footnotesize Aggregating predictions at model-level}
\For {$\tt k \in \mathcal D$}{

$\tt d_{z,k} \gets \text{Behavioral data of } z \text { in component } k$

$\tt P_k \gets$ Specialized predictors in $\tt k$

$\tt \langle R_k, S_k \rangle \gets P_k (d_{z,k})$ \Comment {Get predictions } 

$\tt Expert_k \gets $ Expert set of predictors in $\tt k$  \Comment {(derived from $\tt Prior\-Knowledge$) }

$\tt \langle \hat{r}_k, \hat{c}_k \rangle \gets$  $\tt MA \text{ (}     R_k, S_k, Expert_k$) 

\Comment {Highly confident prediction in $\tt R_k$}

}

\tcc { Aggregating predictions at components Level}

$\tt \hat{R}$ $\gets$ $\tt \{ \hat{r}_{k}$, $\forall \text{ } \tt k$ $\in$ $\tt \mathcal{D}\}$ \Comment Prediction of each component

$\tt \hat{C}$ $\gets$ $\tt \{ \hat{c}_{k}$, $\forall \text{ } \tt k$ $\in$ $\tt \mathcal{D}\}$ \Comment Confidence of each component

$\tt \hat{E} \gets \{ \text{Prior-known }
\text{strength of} \text{ component } k \text{ to  }
\text{predict } \hat{r}_k \} $ \Comment {Derived from $\tt Prior\-Knowledge$}

$\tt L$ $\gets$ Number of processes in the host machine \Comment System load

$\tt \langle \hat{r}_{\mathbb B}, \hat {c}_{\mathbb B} \rangle$ $\gets$ $\tt CA \text{ (} \hat{R}, \hat{C}, \hat{E}, L$)

\Comment {Highly confident prediction among components}

\Return $\tt \langle \hat{r}_{\mathbb B}, \hat {c}_{\mathbb B} \rangle$
}

\caption{SUNDEW}
\label{alg:sundew}
\end{algorithm}

{\flushleft \bf Aggregation.} The independent predictions in $\tt R_k$ are likely to conflict, as shown in Table~\ref{tab:conflict}A, which shows an example output of the specialized predictors in the network component when tested with a backdoor sample. While four predictors (e.g., the ones specialized for banker and backdoor) detect the sample as malware, others predict it as benign. The conflicts arise as each predictor has different estimates for maliciousness ($\tt S_k$), including probability-estimates (Figure~\ref{fig:prob}) or the number of malicious rows in a program (Figure~\ref{fig:rows}). Similarly, the components may differ in their prediction (Refer to Table~\ref{tab:conflict}B), primarily due to the varying noise levels that affect the probability-estimates. While the network component flags the sample as malware, OS and hardware component predicts it as benign. For an unbiased aggregation, while leveraging the benefit of the optimal predictor in each component, \name adopts a two-level aggregation as shown in Figure~\ref{fig:hlv}. Thus, $\tt \mathcal A = \{ MA (), CA()\} $ is a set of aggregator functions that compare the confidence of each predictor to resolve conflicts inside each component and among the components. In either case, the statistic $\tt S_k$ is not sufficient, as aggregating based on the maximum or average of $\tt S_k$ may not relay the optimal prediction (gray cells) to the output (red cells) in most cases (Refer to Table~\ref{tab:conflict}).

\begin{table}[t]
\small

\caption{\small \textbf{(A)} Conflicts among specialized predictors when tested with a backdoor sample in the network component. While four specialized predictors (e.g., banker, backdoor) detect the sample, four others predict it as benign. Naive aggregation schemes based on $\tt S_k$ (probability) may not be  optimal to relay the output of the backdoor-specialized predictor (gray cell) to the output (red cells). Similarly, \textbf{(B)} illustrates the conflict among different components about the sample. While the network component detects it as malware, OS and hardware components flag it as benign.}
\label{tab:conflict}
\vspace{-2mm}

\scriptsize
\begin{tabular}{|M{0.1\linewidth}|M{0.025\linewidth}|M{0.025\linewidth}|M{0.025\linewidth}|M{0.025\linewidth}|M{0.025\linewidth}|M{0.025\linewidth}|M{0.025\linewidth}|M{0.025\linewidth}|M{0.025\linewidth}|M{0.035\linewidth}|M{0.036\linewidth}|}
\hline

\multicolumn{9}{|c|}{\textbf{\begin{tabular}[c]{@{}c@{}} (A) Output from predictors inside a component \end{tabular}}} &

    \multicolumn{3}{c|}{\textbf{Aggregation}} \\ \hline

    \begin{tabular}[c]{@{}l@{}}Speciali-\\zed \\ predictor \end{tabular}
 
 & \rotn{Cryptominer} & \rotn{Banker} & \rotn{Spyware} & \rotn{Backdoor} & \rotn{Ransomware} & \rotn{PUA} & \rotn{Downloader} & \rotn{Deceptor} & \rotn{Majority?} & 

{\rotn{\begin{tabular}[l]{@{}l@{}}Maximum\\ probability \end{tabular}}} & {\rotn{\begin{tabular}[l]{@{}l@{}}Average\\ probability \end{tabular}}} \\ \hline
Prob-1 & 0.47 & 0.38 & 0.33 & 0.73 & 0.33 & 0.34 & 0.19 & 0.49 & - & 0.73 & 0.41  \\
Prob-0 & 0.53 & 0.62 & 0.67 & 0.27 & 0.67 & 0.66 & 0.81 & 0.51 & - & 0.81 & 0.59   \\ \hline
Prediction               & 1                               & 1                          & 0                           & \cellcolor{gray}\textbf{1}                   & 0                              & 0                       & 0                              & 1                            & \cellcolor{red!30}\textbf{?}         & \cellcolor{red!30}\textbf{0}                                      & \cellcolor{red!30}\textbf{0}    \\\hline

\end{tabular}

\vspace{2mm}
\begin{tabular}{|M{0.1\linewidth}|M{0.1\linewidth}|M{0.04\linewidth}|M{0.11\linewidth}|M{0.1\linewidth}|M{0.11\linewidth}|M{0.1\linewidth}|}
\hline

\multicolumn{4}{|c|}{\textbf{\begin{tabular}[c]{@{}c@{}} (B) Output from each component$^*$ \end{tabular}}} &

    \multicolumn{3}{c|}{\textbf{Aggregation}} \\ \hline

& \textbf{Network} & 
\textbf{OS}  & \textbf{Hardware} & \textbf{Majority?} & 
\textbf{Maximum} & \textbf{Average}\\ \hline

Prob-1 & {0.73} & {0.58} & {0.20} & {-} & {0.73} & {0.50}  \\
Prob-0 & {0.27} & 0.42 & {0.80} & {-} & {0.80} & {0.49}   \\ \hline
Prediction & {\cellcolor{gray}\textbf{1}} &     0  & {0}                           & {\cellcolor{red!30}\textbf{0}}         & {\cellcolor{red!30}\textbf{0}}                            & {\cellcolor{red!30}\textbf{0}}    \\\hline      
\end{tabular}

* Assuming each component outputs its best prediction.\\
Prediction of 1 indicates malware class, and 0 indicates benign class. Accordingly, \textbf{Prob-1} indicates the probability of the sample being malware. \textbf{Prob-0} indicates the probability of the sample being benign.
\vspace{-3mm}
\end{table}

{\flushleft \bf Prior knowledge.} To validate the confidence put forward by $\tt S_{k}$, \name also leverages the knowledge built using past experience or domain insights. Thus, $\tt PriorKnowledge(P_{k,j})$ is a comparative measure of prior-known efficiency of predictor $\tt P_{k,j}$ to detect malware $\tt m_j$ in component $\tt k$,  $\tt \forall k \in \mathcal D$, $\tt \forall j \in [1,n]$. An example of such a measure is the F1-Score of $\tt P_{k,j}()$ observed in the train-validate phase or past deployments of \nameA. Based on this measure, some predictors are experts (more confident than others) in a component. For instance, backdoors' operations are known to be network-intensive from domain insights. Thus, if the backdoor-predictor in the network component predicts a program as a backdoor, its prediction is likely to be the most accurate. Accordingly, the expert set $\tt Expert_k$ of a component $\tt k$ (Line 8 in Algorithm~\ref{alg:sundew}) is the set of predictors having high prior-known strengths in $\tt k$, given by,
\begin{equation}
\tt Expert_k = \{ m_j, | \text{ } Prior\-Knowledge (P_{k,j}) > \eta, \text{ } m_j \in \mathcal{M}\} \enspace ,
%
    \label{eqn:experts}
\end{equation}

\noindent for a domain-specific configurable threshold $\eta$.

{\flushleft \bf Model-level Aggregation.} To deduce the most confident prediction from the output of all predictors inside a component $\tt k$ (i.e., $\tt R_k = \{r_{k,j}, \forall j \in [1,n] \}$, per Equations \ref{eqn:pkj} and \ref{eqn:rkj}), \name invokes the function $\tt MA ()$ with inputs: the predictions ($\tt R_k$), statistics ($\tt S_k = \{ s_{k,j}, \forall \tt j \in \tt [1,n] \}$), and expert set ($\tt Expert_k$) of predictors in $\tt k$ (Line 9). Internally, $\tt MA ()$ uses $\tt S_k$ and $\tt Expert_k$ to evaluate the confidence of each prediction in $\tt R_k$, and return $\tt \hat{r}_k$ and $\tt \hat{c}_k$, the most confident prediction of the component and its confidence measure (Line 9).

{\flushleft \bf Component-level aggregation.} 
At the components level, \name has three independent predictions $\tt \hat{R}$ from each $\tt MA ()$ in Line 9, wherein $\tt \hat{R} =  \{\hat{r}_{k}$, $\forall \text{ } \tt k \in \mathcal{D}\}$ (Line $11$). Similarly, \name has their corresponding confidence values  $\tt \hat{C} = \tt \{\hat{c}_{k}$, $\forall \text{ } \tt k \in \mathcal{D}\}$ (Line $12$). As the predictions in $\tt \hat{R}$ are likely to differ, \name again leverages the prior known strengths of the specialized predictor for the predicted malware class $\tt \hat{r}_{k} \in \hat{R}$ in component $\tt k$, given by,

\begin{equation}
\tt \hat{E} = \{ \tt \text{ } Prior\-Knowledge (P_{k,\hat{r}_k}) \text{ } \forall k  \in \mathcal{D} \} \enspace.
    \label{eqn:ehat}
\end{equation}

\noindent These scores are indicative of the confidence of a component $\tt k$ in predicting $\tt \hat{r}_{k}$. As noise induced by system load could affect detection, \name also considers the system load, $\tt L$. For example, $\tt L$ can be the number of processes executing at the host machine (Line $14$). Finally, \name invokes the component-aggregator $\tt CA ()$ to aggregate the three predictions. The function $\tt CA()$ takes as input the predictions ($\tt \hat{R}$), confidences ($\tt \hat{C}$), and prior-known strengths ($\tt \hat{E}$), and the system load ($\tt L$). Similar to $\tt MA()$, it evaluates the confidence of each component, and outputs the highly confident prediction ($\tt \hat{r}_{\mathbb B}$) and its confidence ($\tt \hat{c}_{\mathbb B}$), as the final output of \name (Line $15$). In the next section, we discuss how \name builds insights from predictor statistics and prior-known strengths for case-sensitive detection of malware.

\section{Insightful Aggregation of Predictions}
\label{sec:aggregation}

The functionalities of the model and component aggregators are different. A model aggregator ($\tt MA()$) resolves conflicts among predictors that use the same data source to test a sample, but have different definitions of the boundary between positive and negative classes. In contrast, a component aggregator ($\tt CA()$) resolves conflicts between predictors that use different data sources (having varying levels of noise) for the same sample. While the predictions can be aggregated in different ways, the optimal aggregation mechanism for different malware classes and components varies due to the differences in the boundary definitions and the noise levels. For instance, naive comparisons such as majority~\cite{salo2019dimensionality,li2021framework} or averaging of statistics~\cite{sayadi2018ensemble,zhou2020building}) may not lead to the optimal aggregation for all malware classes. Thus, \name leverages a configurable $\tt MA()$ and $\tt CA()$ that explore different mechanisms on test-time predictor statistics, prior-known strengths, and system load to relay the optimal prediction as the final output, as discussed next.

\subsection{Model-Aggregator}
Algorithm~\ref{alg:model} describes the model-aggregator $\tt MA ()$ function. For a given component $\tt k$, it takes as input the predictions ($\tt R_k$), corresponding statistics ($\tt S_k$), and expert set of predictors $\tt Expert_k$. As each predictor has different definitions of malware-benign boundary (Figure~\ref{fig:prob}), $\tt MA ()$ aggregates predictions in a two-step process, by achieving consensus first on the maliciousness of the program and then on the specific class of the malware.

{\flushleft \bf Binary consensus on maliciousness.}
Firstly, $\tt MA()$ assesses the predictions and statistics to vote if the test program is malware or benign using a function $\tt ConsensusIfMalware()$(Line 2 of Algorithm~\ref{alg:model}). Multiple alternatives are possible for realizing $\tt ConsensusIfMalware()$, including consensus based on logical-OR, majority-vote, confidence, or learning. Naive mechanisms infer malware if at-least one of the predictions ({\em logical-OR}), or most predictions are malware ({\em majority-vote}). However, such approaches can significantly increase the false positives. 

A {\em most-confident} $\tt ConsensusIfMalware()$ infers malware if the aggregated confidence of malware predictions is higher than benign predictions. It aggregates confidences using the {\em mean-probability-difference}, which is the mean difference between the probability-estimates of malware (Prob-1 in Table~\ref{tab:conflict}) and benign class (Prob-0) of all predictors. It infers malware when mean-probability-difference $>0$, and benign otherwise. 
Another potential metric to aggregate confidences is  {\em mean-maliciousness-difference}, which is the mean difference between the percentage of malicious and benign rows inferred for a program by all predictors. However, we find that mean-maliciousness-difference is sub-optimal for $\tt ConsensusIfMalware()$ as the percentage of malicious rows varies for each class (See Figure~\ref{fig:rows}).

Alternatively, a learning-based $\tt ConsensusIfMalware()$ uses trained models to infer malware. Two configurations are possible for such models. A {\em booster} learns to minimize the loss function of the specialized predictors. On the other hand, a {\em multiplexer} learns to multiplex the output of the specialized predictor to the final output of \nameA. Both these configurations train their models with the predictor outputs (Refer to Equation~\ref{eqn:pkj}) observed for all programs $\tt z \in$ $\tt Z_{train}$, which is the set of programs in the training phase. Thus, these models train on $\tt X_k = [\{\tt P_{k,j} (d_{z,k}), \forall j \in \mathcal M, \forall z \in Z_{train}]$ to predict target labels $\tt Y_k = [y_{z,k} | y_{z,k} \in \mathcal M, \forall z \in Z_{train}]$. The target label $\tt y_{z,k}$ for a program $\tt z$ is different for booster and multiplexer. As the booster minimizes the loss function of the component, its $\tt y_{z,k}$ is the actual-class $\tt c \in \mathcal M$ of the program $\tt z$. On the other hand, the multiplexer aims to relay the best-case prediction of $\tt z$ to the output. Thus, its $\tt y_{z,k}$ is the predicted class of $\tt z$ when tested on the predictor specialized for class $\tt c$ in component $\tt k$.

\begin{algorithm}[!t]
\small
\DontPrintSemicolon

\KwIn{$\tt R_k = \{r_{k,j} \text{, Predictions from } \tt P_{k,j} \forall \tt j \in \tt [1,n] \}$, $\tt S_k = \{s_{k,j} \text{, Statistics from } \tt P_{k,j} \forall j \in [1,n]\}$, {$\tt Expert_k$:~Expert set of component $\tt k$}.
}

\KwResult{$\tt \langle \hat{r} ,  \hat {c} \rangle$: Final vote and confidence.}
    
\Begin{

    \tcp{Check if predictions in $\tt R$ indicate malware}
    ($\tt label$, $\tt probmalware$) 
    $\gets$ $\tt ConsensusIfMalware(R_k, S_k$) 
    \textit{}

     \If {$\tt label \text{ is }$ $\tt malware$ } {

    \tcp{Identify the set of most confident predictor(s) from $\tt R$ | $\tt r_{k,j}$ is $\tt 1$.}
    $\tt C\_Set$ $\gets$ GetConfidentSet ($\tt R_k, S_k, Expert_k$)

  $\tt \hat{r} \gets \text{Malware class of the most risky in } C\_Set$
   
  $\tt \hat{c}$ $\gets$ $\text{Probability estimate of } \tt \hat{r}$}
   
    \Else{
  $\tt \langle \hat{r}, \hat {c} \rangle \gets \langle benign$ , $\tt 1 - probmalware) \rangle$}
  \Return $\tt \langle \hat{r}, \hat{c} \rangle $
}    
    
\caption{Model-Aggregator (for component $\tt k$)}
\label{alg:model}
\end{algorithm}

\begin{table}[t]
\centering
\caption{\footnotesize \textbf{Aggregation-loss with different alternatives for} $\tt MA ()$ \textbf{and} $\tt CA ()$. The comparison baseline is the F1-Score of the specialized predictor that is optimum for the program. An aggregation-loss of zero indicates that the $\tt MA()$ and $\tt CA()$ can relay the prediction of the optimum specialized predictor to the final output for any program. A negative loss indicates that the aggregation is able to improve the detection performance beyond that of the specialized predictor.}

\vspace{-4mm}
\label{tab:gerror}
\begin{tabular}{M{0.01\linewidth}M{0.04\linewidth}M{0.3\linewidth}
M{0.12\linewidth}M{0.12\linewidth}M{0.12\linewidth}}

\addlinespace[0.5mm]
\toprule
\addlinespace[0.5mm]
& \textbf{} &  & \textbf{Network}  & \textbf{OS} & \textbf{Hardware} \\ \cline{3-6} 
& & \textit{Baseline}           & $0\%$     & $0\%$         & $0\%$     \\ 
& & \textit{Logical-OR}         & $41.92\%$    & $80.05\%$     & $80.05\%$    \\ 
& & \textit{Majority-vote}      & $33.66\%$    & $6.81\%$     & $68.82\%$    \\ 
& & \textit{Most-Confident}     & $59.04\%$ & $7.12\%$      & $7.12\%$    \\ 
& & \textit{Multiplexer}        & $5.18\%$  & $4.64\%$     & $4.64\%$    \\ 


& \multirow{-7}{*}{\textbf{\rot{\begin{tabular}[c]{@{}c@{}}\textbf{Binary}\\\scriptsize $\tt Consensus$\\ $ \tt IfMalware() $\end{tabular}}}} & 

\textit{Booster}    &
\cellcolor{green!20} $4.76\%$  & \cellcolor{green!20}$-0.28\%$ & \cellcolor{green!20}$-1.35\%$ \\ \addlinespace[1mm] \cline{2-6} \addlinespace[0.5mm]

& & {\color[HTML]{333333} }               & \textbf{Network}  & \textbf{OS} & \textbf{Hardware} \\ 
\cline{3-6} 
& & \textit{Baseline}       & $0\%$     & $0\%$         & $0\%$            \\ 
& & \textit{Confidence}     & $4.75\%$  & $21.51\%$     & $0.56\%$            \\ 
& & \textit{Prio-knowledge} & $34.16\%$  & \cellcolor{green!20}$-1.74\%$     & $0.64\%$            \\ 
\multirow{-13}{*}{\textbf{\rot{$\tt MA ()$}}} & \multirow{-6}{*}{\textbf{\rot{\begin{tabular}[c]{@{}c@{}}\textbf{Multi-Class}\\\scriptsize {$\tt GetConfidentSet()$}\end{tabular}}}} &

\textit{\begin{tabular}[c]{@{}c@{}}Confidence-Window\end{tabular}} & \cellcolor{green!20}$4.79\%$          & $21.51\%$      & \cellcolor{green!20}$-2.18\%$            \\ \addlinespace[0.5mm] \hline 

\multicolumn{2}{l}{}           &               & \textbf{Binary}  &  \multicolumn{2}{c}{}  \textbf{Multi-Class }        \\ 
\cline{3-6}

\multicolumn{2}{l}{}           & \textit{Most-Confident}               & $-1.42\%$  &    \multicolumn{2}{c}{} $12\%$          \\ 
\multicolumn{2}{l}{\multirow{-2}{*}{\textbf{\rot{$\tt CA ()$}}}}             & \textit{Prior-Knowledge}              &   \cellcolor{green!20}$ -1.42\%$ & \multicolumn{2}{c}{} \cellcolor{green!20}$7.86\%$           \\ \bottomrule
\end{tabular}
\vspace{-0.2cm}
\end{table}

Table~\ref{tab:gerror} enlists the {\em aggregation-loss} of $\tt MA()$, which is the difference between the F1-Score of $\tt ConsensusIfMalware()$ as compared to the baseline, i.e., the F1-Score achieved by a specialized predictor that is optimum for the program. An aggregation-loss of zero indicates that $\tt MA()$ is able to relay the inference of the optimum specialized predictor to the output of \nameA. On the other hand, negative loss indicates that $\tt MA ()$ can boost the detection performance beyond that of the specialized predictor by minimizing its loss function. Naive voting mechanisms such as logical-OR or majority-vote have high aggregation-losses. While the logical-OR function leads to high false positives, majority-vote fails when less than half of the predictors can detect the malware sample. Similarly, the performance of the most-confident $\tt ConsensusIfMalware()$  can be sub-optimal as the range of probability estimates that differentiates malware and benign are different for each specialized predictor (refer to Figure~\ref{fig:prob}). In contrast, the learning-based mechanisms can learn these class-specific probability estimates effectively to reduce aggregation losses. Specifically, the booster improves the performance of both OS and hardware components by at least 1\% while reducing the aggregation-loss to as low as $4.76\%$ in the network component.

\begin{figure}[t]
    \centering
    {\includegraphics[width=\linewidth]{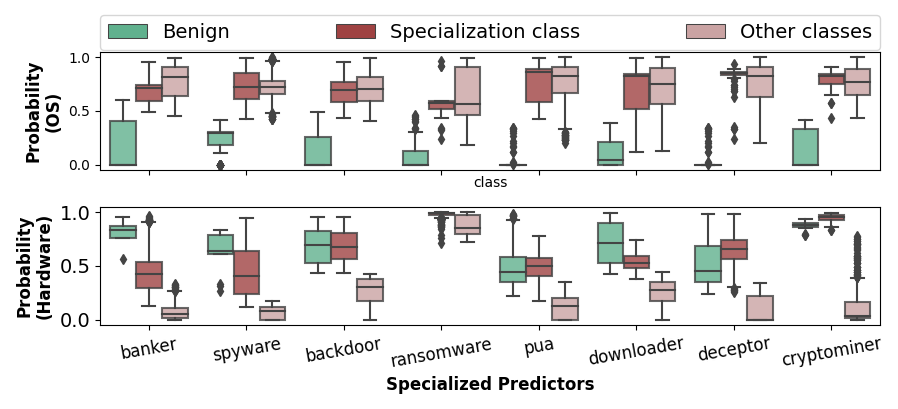} }
    \vspace{-0.7cm}
    \caption{\footnotesize Probability distribution of different specialized predictors in the OS and hardware component, when tested with data from the benign class (green), their corresponding specialization (red), or any other class (light red). The distributions overlap significantly in OS as compared to that in the network (Figure~\ref{fig:prob}) and hardware, making confidence (probability-estimate), a poor metric to aggregate class in $\tt MA()$.\vspace{-0.3cm}}
    \label{fig:backdoor}
\end{figure}

{\flushleft \bf Multi-class consensus.} After consensus on the maliciousness of the program, the probability estimates ($\tt S_k$) of individual predictors could help identify the most {\em confident} predictor, and hence the class. However, these estimates get unreliable when a specialized predictor attempts to predict on data of any other class (light-red boxes in Figure~\ref{fig:prob}). The figure plots these estimates of different specialized predictors in the network component. Figure~\ref{fig:backdoor} plots the corresponding distributions in the OS and hardware component when tested with data of benign class (green), the corresponding specialization (red), or from any other class (light red). The overlapping distributions especially in the OS and hardware component makes identifying the malware objective class non-trivial. Alternatively, some predictors have a {\em confidence-window}, wherein the aforementioned distributions (distance between their inter-quartile range) are far apart (e.g., spyware in network component in Figure~\ref{fig:prob}). If the test-time statistics of a predictor fall in such confidence windows, the predictor can be considered highly confident. As evident, confidence windows are beneficial in the network (Figure~\ref{fig:prob}) or hardware component (Figure~\ref{fig:backdoor}), whereas in OS, they can get unreliable. An alternative is to employ prior-knowledge and prioritize predictions of the {\em expert-set} alone (refer Equation~\ref{eqn:experts}).

\begin{figure}[t!]
    \centering
    {\includegraphics[width=0.8\linewidth]{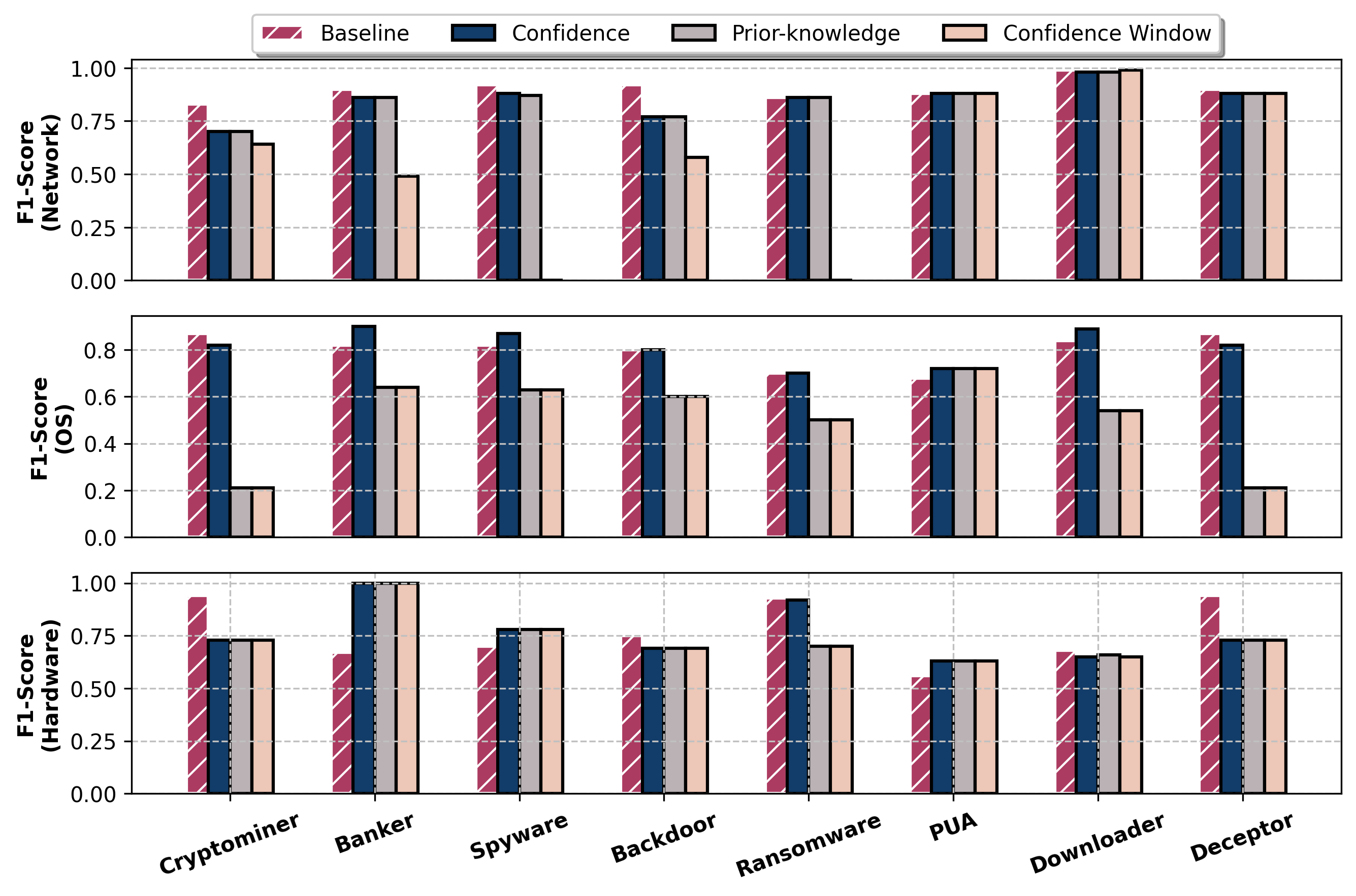} }
    \vspace{-0.4cm}
    \caption{ \small Detection F1-Score observed for different classes for different alternatives for MA() at network, OS, and hardware. The baseline for comparison is the F1-Score achieved with the specialized predictor optimum for each class. }
    \vspace{-0.2cm}
        \label{fig:gsummary}
        
\end{figure}

Accordingly, if the test program is a malware, $\tt MA()$ evaluates the confidences of all predictions $\tt r_{k,j} \in R_k$ which predicted malware (i.e. $\tt r_{k,j} = 1$ in Equations~\ref{eqn:pkj} and~\ref{eqn:rkj}). For this, it employs a configurable function $\tt GetConfidentSet ()$ that evaluates the statistics ($\tt S_k$) and expert set ($\tt Expert_k$) to return a confident set of predictor(s), $\tt C\_Set$ (Line 4 of Algorithm~\ref{alg:model}). To compute such a set, $\tt GetConfidentSet ()$ can use one of the following metrics: \textbf{(1)} {\em confidence}, that returns the predictor with high probability-estimates; \textbf{(2)} {\em prior-knowledge}, that prioritizes expert predictors in $\tt Expert_k$ to choose predictors with high probability-estimates; or \textbf{(3)} {\em confidence-window}, that returns the predictors whose statistics fall within their respective confidence window. These options can return a set of confident predictors. To resolve the contention in $\tt C\_Set$ in such cases,  $\tt MA()$ prioritizes the classes in accordance to the risk categories~\cite{risklevel} and outputs the most risky class in $\tt C\_Set$ as the prediction of the component (Line 6).

Figure~\ref{fig:gsummary} evaluates the F1-Score of $\tt MA ()$ when tested with any program, for different alternatives of $\tt GetConfidentSet()$, against the baseline F1-Score achieved with the corresponding specialized-predictor that is optimum for the program. The detection is inferred as correct if the risk level of the predicted class is the same or higher than that of the program.  $\tt MA()$ can restrict the aggregation-losses to as low as $4\%$ at the component outputs to aggregate the class for {\em any} program (Refer to Multi-class row in Table~\ref{tab:gerror}). While the confidence metric is the most effective for the network component, prior-knowledge and confidence-window metrics are effective for the OS and hardware, respectively.

\subsection{Component Aggregator}

\begin{algorithm}[!t]
\DontPrintSemicolon
\small
\KwIn{$\tt \hat{R} = \{ \hat{r}_{N},\hat{r}_{O}, \hat{r}_{H}\}\text{:Predictions from }MA () \text{for all } k \in \mathcal D$, 
$\tt \hat{C} = \{ \hat{c}_N,\hat{c}_O, \hat{c}_H\}\text{:Confidences from } MA () \text{ for all } k\in\mathcal D$, 
$\tt\hat{E} \text{: Prior-knowledge}$, $\tt L$: System Load (Number of processes in the system).
}
\KwResult{$\tt \langle \hat{r}_{\mathbb B} , \hat{c}_{\mathbb B} \rangle$: Final label and confidence.}
\Begin{

 \If {$\tt L <  \tau $ } {

    \tcp{At low system loads, all components are reliable.}
$\tt C\_Set$ $\gets$ $\tt \text{Compute confident set} (\hat{R}, \hat{C}, \hat{E})$ 

 $\tt \hat{r}_{\mathbb B} \gets \text{Class of the most risky among } C\_Set$ 
 
 $\tt \hat{c}_{\mathbb B} \gets \text{Confidence of the most risky among } C\_Set$
  }
\Else{
\tcp{At higher system loads, OS is the most reliable. }
   $\tt \langle \hat{r}, \hat {c} \rangle \gets \langle \hat{r}_O , \hat{c}_O \rangle$
}    

\Return $\tt \langle \hat{r}_{\mathbb B}, \hat{c}_{\mathbb B} \rangle$
}
\caption{Component-Aggregator}
\label{alg:f}

\end{algorithm}


 The $\tt CA()$ is responsible for choosing the most confident prediction in $\tt \hat{R} =  \{\hat{r}_{k}$, $\forall \text{ } \tt k \in \mathcal{D}\}$, where $\tt \hat{r}_{k}$ is the output prediction aggregated by the $\tt MA()$ in each component $\tt k$ (Line 9 of Algorithm~\ref{alg:model}). $\tt CA ()$ can weigh components based on the empirical confidence of their prediction observed at test-time ($\tt \hat{C} = \tt \{\hat{c}_{k}$, $\forall \text{ } \tt k \in \mathcal{D}\}$ (Line 9 of Algorithm~\ref{alg:model})),  
 or their prior-known strengths ($\tt \hat{E}$, Equation~\ref{eqn:ehat}) in predicting $\tt \hat{R}$. Alternatively, it can weigh components based on their resilience to noise and system load. At higher loads, the OS component is the most stable and noise-free, as OS logs are collected specifically to the process PID. In contrast, the network and hardware can get noisier with an increase in the number of processes.

 Accordingly, $\tt CA()$ (presented in Algorithm~\ref{alg:f}) takes as input the predictions from each component ($\tt \hat{R}$), corresponding confidences ($\tt \hat{C}$), prior-knowledge ($\tt \hat{E}$), and the system load ($\tt L$) which is the number of processes in the system. At lower system loads (Line 2), it computes a confident set $\tt C\_Set$ using multiple options such as : 
 \textbf{(1)} {\em most-confident} selects the prediction which has high $\tt \hat{c}$;  \textbf{(2)} {\em prior-known} selects the prediction which has high prior-knowledge scores; or, \textbf{(3)} {\em majority} selects the prediction that is common between at least two components (Line 3). Similar to $\tt MA ()$, $\tt CA()$ resolves contentions in $\tt C\_Set$ by choosing the most risky class in $\tt C\_Set$ as the final aggregated prediction (Lines 4 and 5). On the other hand, at higher system loads, the OS component is the most stable, and hence $\tt CA()$ outputs predictions of the OS component directly (Line 7).

 \begin{figure}[t!]
    \centering
        \vspace{-4mm}
    {\includegraphics[width=\linewidth]{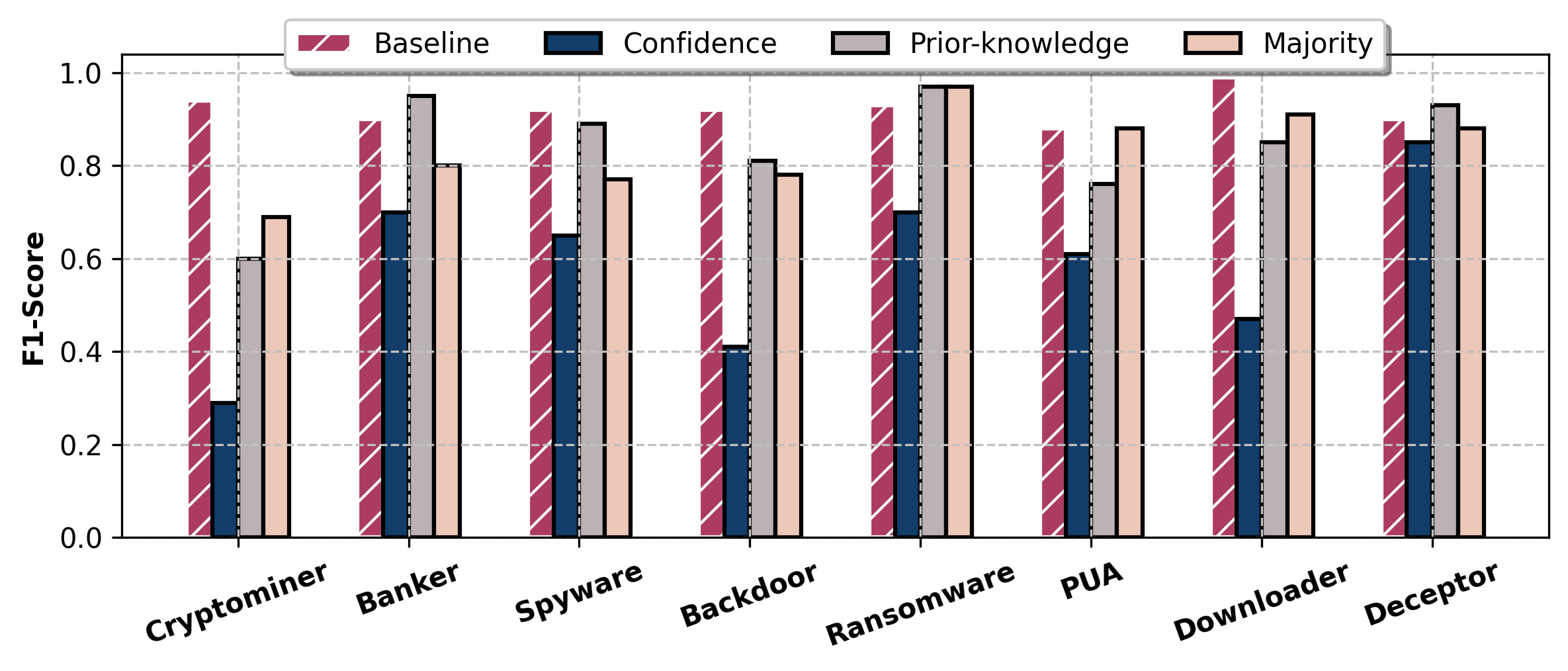} }
    \vspace{-5mm}
    \caption{\small Detection F1-Scores for different classes using different confident computing functions in $\tt CA()$. Employing prior-knowledge gives the best F1-Score for most malware classes.}
        \label{fig:fsummary}

        \vspace{-3mm}
\end{figure}

Figure~\ref{fig:fsummary} evaluates the F1-Score obtained for different alternatives of $\tt CA()$ against the performance of the component that is optimum for each class. Given any program, we consider the prediction to be correct if the risk of the predicted class is the same or higher than that of the program class. Exploiting prior-knowledge, $\tt CA ()$ is able to detect any malware boosting the performance beyond that of its best-case specialized predictor by at least $1.42$\%, and detect the objective class of the program with a loss as low as $7.86\%$ (Refer to the row $\tt CA ()$ in Table~\ref{tab:gerror}).

\vspace{-0.3cm}
\section{Implementation and Evaluation}
\label{sec:implementation}
\vspace{-0.1cm}
In this section, we discuss the real-world behavioral data used to build the ensemble, followed by the implementation and evaluation of \nameA. 
\vspace{-0.25cm}
\subsection{Real-World Behavioral Dataset}
For unbiased cross-dimensional analyses, \name requires access to a simultaneous capture of network, OS, and hardware run-time trails of a large corpus of malware samples of different classes. \name relies on the RaDaR dataset~\cite{radar} that provides such a comprehensive view of the real-world activity across the system stack of diverse Windows malware families labeled with their attack objective. RaDaR is collected by executing live malware samples (2017 ongoing) on a real-world testbed~\cite{jugaad} with Internet connectivity, in a timely manner, when their remote command-and-control servers are highly likely to be active. Each sample is executed for 2 minutes in an automated manner, which is known to be sufficient to elicit malicious activities of most malware samples~\cite{alexander:2021:second}. For a fair comparison, the benign samples are executed in an automated manner similar to malware, as user interactions are easily distinguishable, unlike the stealthy malware activities.

The dataset~\cite{radar} provides a comprehensive set of popular features extracted based on prior works~\cite{Bartos:2016:invariantrep, guofei:2008:botminer, perdisci:2010:behavioral,Bayer:2009:scalableMalwareClustering,Canali:2012:call-based-malware-detection,das:2016:sematicsonline, sayadi2018ensemble}, from $7$ million network packets, $11.3$ million OS system call traces, and $3.3$ million hardware events collected for $10,434$ samples. These features include $58$ features at network, $11$ at OS, and $54$ micro-architectural events at the hardware~\cite{radar}. Each row in the data represents a snapshot of network flow\footnote{All communications having the same source and destination IP address, and source and destination port belong to a flow. Thus the network packets are grouped into traffic flow summaries}, system call in OS, and periodic HPC measurement in 100ms intervals in hardware. Table~\ref{tab:dataset} summarizes the number of snapshots corresponding to each malware class from the three data-sources. With data of 10,434 samples evenly spread across 30 well-known malware families belonging to 8 different classes (attack objectives) and benign applications, RaDaR~\cite{radar} provides a diverse representation of malware classes for evaluations.

{\flushleft \bf Train-validate-test partitions.} Finally, we split the dataset in a 70:15:15 ratio into the train, validate and test sets. Specifically, we ensure that the train set does not contain samples, whose data is collected at a later point of time than a sample in validate/test sets to prevent experimental biases~\cite{feargus:2019:tesseract}. To ensure unbiased learning, the train set contains an even distribution of benign and malware classes.

\begin{table}[t]
    \centering
    \caption{\small  Summary of behavioral snapshots of different malware classes from the three data-sources in RaDaR~\cite{radar}. Snapshots indicate the number of flows in the network, system call traces in OS, and periodic HPC logs in the hardware.}
    
    \vspace{-0.2cm} 
\begin{tabular}{|m{0.1\linewidth}|m{0.045\linewidth}|m{0.045\linewidth}|m{0.045\linewidth}|m{0.045\linewidth}|m{0.045\linewidth}|m{0.045\linewidth}|m{0.045\linewidth}|m{0.045\linewidth}| m{0.045\linewidth}|}
\hline

\scriptsize

 & \rotn{\textbf{Cryptominer}} & \rotn{\textbf{Banker}} & \rotn{\textbf{Spyware}} & \rotn{\textbf{Backdoor}} & \rotn{\textbf{Ransomware}} & \rotn{\textbf{PUA}} & \rotn{\textbf{Downloader}} & \rotn{\textbf{Deceptor}} & \rotn{\textbf{Benign}} \\ \hline
\textbf{Network} & 992 & 4878 & 11588 & 7845 & 2239 & 7152 & 9277 & 4617 & 8964   \\
\hline
\textbf{OS} & 293K & 772K & 1.9M & 1.5M& 807K & 2M & 1.7M & 440K & 1.9M    \\ \hline
\textbf{Hardware}               & 158K                               & 51K  & 59K  & 371K & 182K                              & 914K  & 502K  & 478K & 578k  \\\hline

\end{tabular}
\vspace{-0.2cm}
    \label{tab:dataset}
\end{table}

\subsection{\name Implementation}
We implement the specialized predictors in Python v3.6.2 using XGBoost\footnote{https://xgboost.readthedocs.io/en/stable/python/} v1.4.2 library. We train each specialized predictor with the train-validate set containing data of the specific malware class and benign programs. Next, we test every specialized predictor with the train-validate sets of all other malware classes to generate conflicting predictions. The resultant predictions and statistics form the train-validate sets for the aggregators. We implement the aggregators using Python LightGBM library v3.3.1\footnote{https://lightgbm.readthedocs.io/en/latest/}. Finally, we evaluate the performance of \name using the test set of malware samples.

\begin{figure}[t!]
    \centering
  \includegraphics[width=\linewidth]{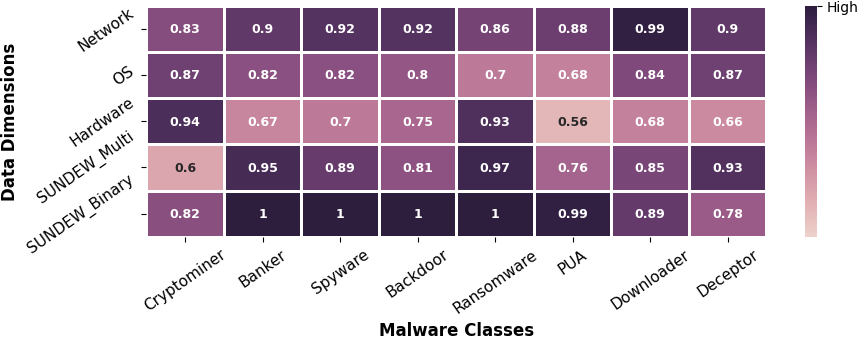} 
\vspace{-0.6cm}
  \caption{\footnotesize The F1-Score observed with \nameA\_Binary and \nameA\_Multi, in comparison with the specialized predictors fine-tuned for each class in each component. \nameA\_Binary infers if a program is malware/benign, whereas \nameA\_Multi infers the class of the program. \nameA\_Binary achieves an F1-Score of 1 for most classes as compared to their best-case specialized predictors in network, OS, or hardware.}
    \label{fig:sundewresults}
\vspace{-0.3cm}
\end{figure}

\vspace{-0.3cm}
\subsection{Evaluation}
We compare the performance and resilience of \name against the best-case specialized predictors of all malware classes, as well as the state-of-the-art malware classifiers. Finally, we evaluate the overheads incurred by \nameA.

{\flushleft \bf Specialized predictors.} Figure~\ref{fig:sundewresults} compares the performance of \name to detect a malware class against the corresponding predictor specialized for that class (and hence the optimum) in different components (data sources). We consider two configurations: \nameA\_Binary measures the F1-Score of detecting if a test sample is malware/benign, whereas \nameA\_Multi measures the F1-Score of inferring the class of the sample. As evident, \nameA\_Binary, with its holistic view of malware activity from the three data sources, an ensemble of specialized predictors, and aggregation, can achieve performance similar to the corresponding specialized predictor for any malware class. The aggregation in \nameA\_Binary boosts the average detection performance beyond that of the best-case specialized predictors in any of the three components, by $1.14\%$. \nameA\_Binary has an F1-Score of $1$ for most malware classes and an average score of $0.93$ for any malware class. While gaining performance in high-risk malware, the performance of low-risk malware slightly drops due to aggregation. On the other hand, aggregating the correct objective class of the sample in \nameA\_Multi incurs an aggregation loss of $7.86\%$. This is because the evaluation considers a detection successful only if the ensemble predicts the actual class or a riskier class for the test sample. Hence, though PUA is successfully detected, its measure in \nameA\_Multi drops as the riskier class Deceptor is chosen when resolving conflicting predictions during aggregation.

\begin{figure}[!t]
\small 
\centering
\captionsetup[subfloat]{}
\subfloat[\footnotesize F1-Scores of detection models based on network and OS under various load conditions for their best detectable classes. The OS-based models are resilient to infiltrating noise from increasing system load.]{%
  \includegraphics[width=0.85\linewidth]{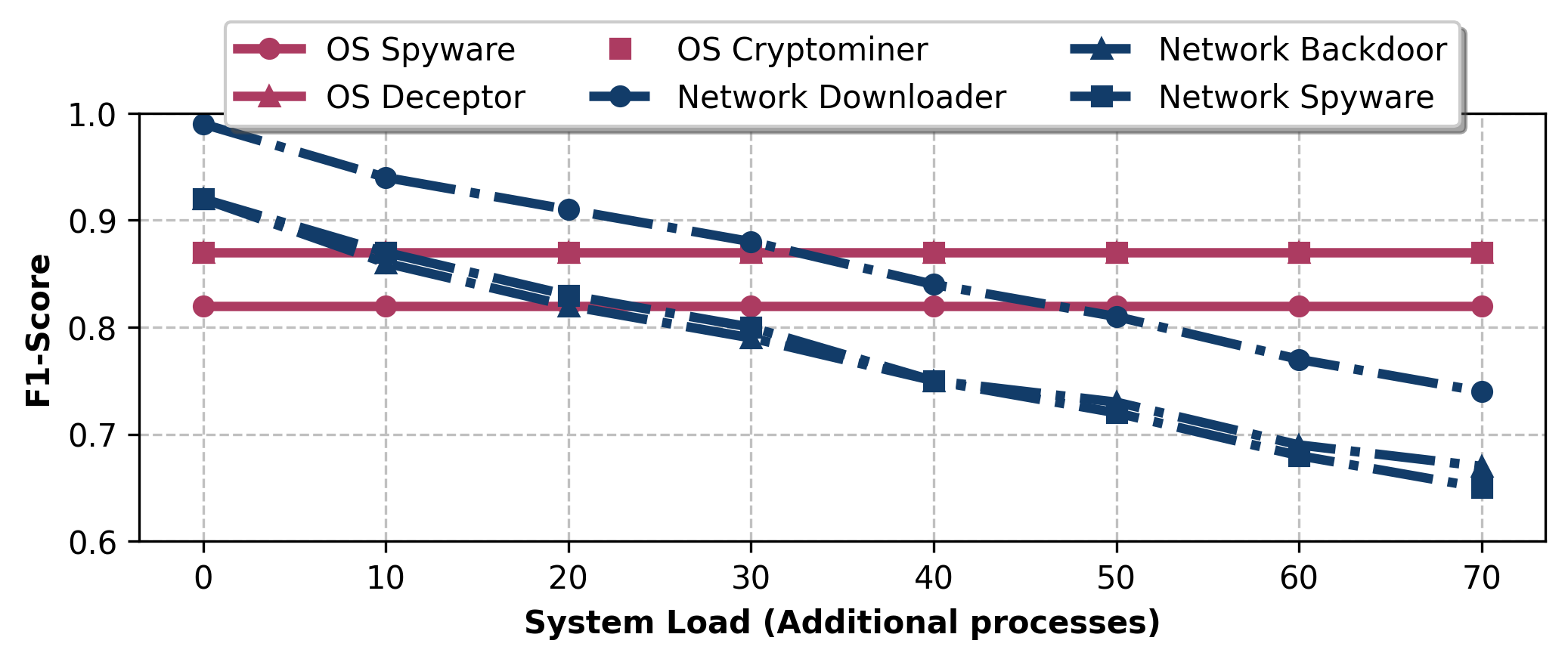}\vspace{-0.3cm}\label{fig:f1score_vs_load_os_nw}}
  \\
\subfloat[\footnotesize F1-Scores of detection models based on network, OS, or hardware and that of \name under varying load conditions for Cryptominer class. \name is able to leverage the best of the three components for detection accuracy while benefiting from the resilience of the OS component.]{
  \includegraphics[width=0.85\linewidth]{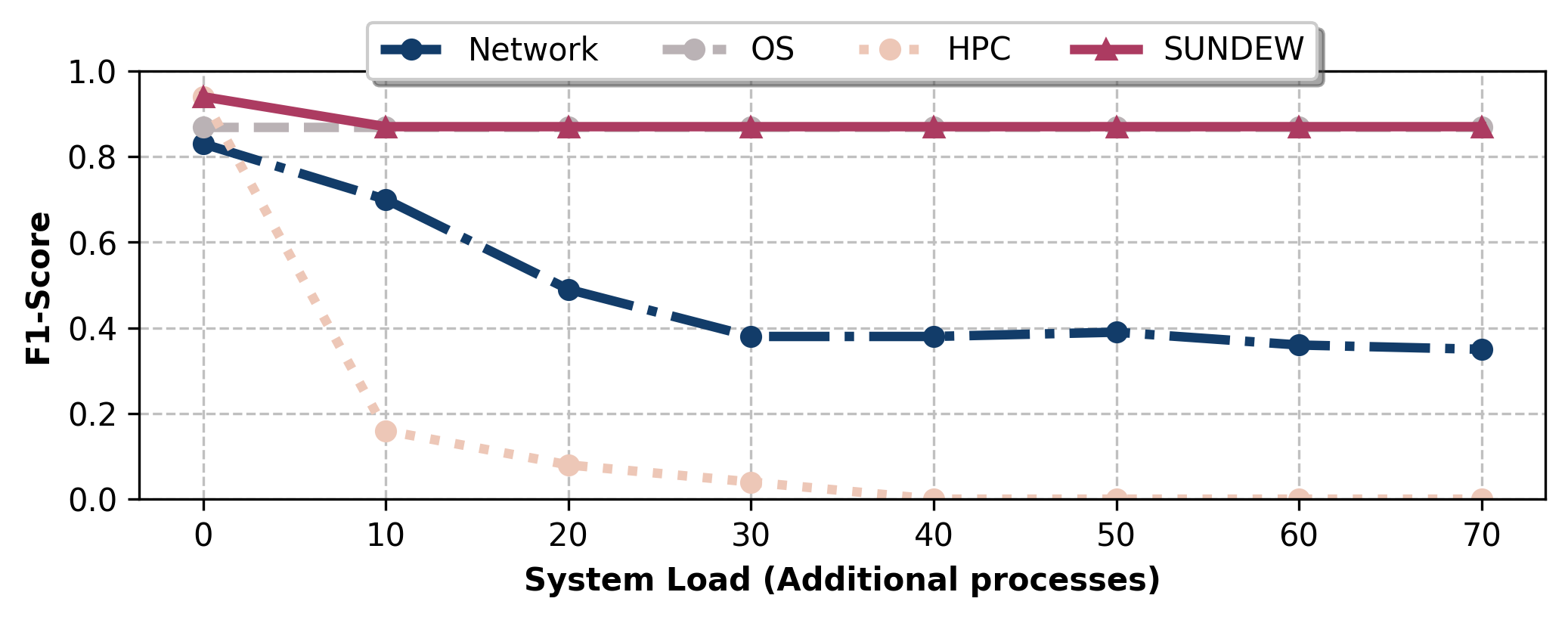}\vspace{-0.3cm} 
\label{fig:f1score_vs_load}}
\vspace{-0.2cm}
\caption{\small Impact of system load on detection efficacy.}
\vspace{-0.4cm}
\end{figure}

{\flushleft \bf Resilience to noise}. We next evaluate \name under varying noise infiltration induced by system load. We use the number of processes in the system to measure noise. To generate data for the experiment, we run benign applications from CNET~\cite{cnet} in multiples of $10$ in the background while running the malware programs and collect the corresponding data at network, OS, and hardware. While these benign applications represent use-case scenarios, an extensive characterization covering a wide range of system loads is planned for future work.  Figure~\ref{fig:f1score_vs_load_os_nw} plots the F1-Score of specialized predictors in network and OS under varying system load conditions. As evident, the performance of the network component decrease, while the OS component is agnostic to system load. 

\begin{table}[t]
    \centering
    \scriptsize
    \caption { \footnotesize Comparison of \name with prior state-of-the-art solutions including single classifiers~\cite{Bartos:2016:invariantrep,das:2016:sematicsonline, sayadi2018ensemble} and single-input ensembles~\cite{tanmoy:2020:ec2} based on \textbf{(A)} F1-Score, and \textbf{(B)} False-positive rate, of detection observed on the RaDaR dataset.}
        \vspace{-0.2cm}
        \label{sundew:tab:priorartf1score}
    \begin{tabular}{|M{0.14\linewidth}|M{0.03\linewidth}|M{0.03\linewidth}|M{0.03\linewidth}|M{0.03\linewidth} | M{0.04\linewidth}| M{0.03\linewidth}|M{0.03\linewidth}|M{0.03\linewidth}|M{0.03\linewidth}| M{0.04\linewidth}|}
    
        \hline
        
        & \multicolumn{5}{c|}{\textbf{\scriptsize\begin{tabular}[c]{@{}c@{}}Detection Performance\\ (F1-Score)\end{tabular}}} & \multicolumn{5}{c|}{\textbf{\scriptsize\begin{tabular}[c]{@{}c@{}}(B)False-Positive Rate\end{tabular}}}\\
        \hline
          
         \multicolumn{1}{|c|}{\textbf{\scriptsize\begin{tabular}[c]{@{}c@{}}Detection\\ model\end{tabular}}} & \multicolumn{3}{c|}{\textbf{\begin{tabular}[c]{@{}c@{}}Single\\ Classifier\end{tabular}}} & \multicolumn{1}{c|}{\textbf{\rot{\begin{tabular}[c]{@{}c@{}}SIE$^*$\end{tabular}}}} & \multicolumn{1}{c|}{\textbf{\scriptsize \rot{\begin{tabular}[c]{@{}c@{}}SUNDEW\end{tabular}}}} 

         & \multicolumn{3}{c|}{\textbf{\begin{tabular}[c]{@{}c@{}}Single \\ classifier\end{tabular}}} & \multicolumn{1}{c|}{\textbf{\rot{\begin{tabular}[c]{@{}c@{}}SIE$^*$\end{tabular}}}} & \multicolumn{1}{c|}{\textbf{\rot{\begin{tabular}[c]{@{}c@{}}SUNDEW\end{tabular}}}}  \\\hline

\multirow{4}{*}{\textbf{\begin{tabular}[c]{@{}c@{}}Component\\$\Rightarrow$\\ Class\  $\Downarrow$\end{tabular}}}

& \rot{{\begin{tabular}[c]{@{}c@{}}N\cite{Bartos:2016:invariantrep} \end{tabular}}}
       
        & \rot{{\begin{tabular}[c]{@{}c@{}}O \cite{das:2016:sematicsonline} \end{tabular}}}    & \rot{{\begin{tabular}[c]{@{}c@{}}H\cite{sayadi2018ensemble} \end{tabular}}}  & \rot{{\begin{tabular}[c]{@{}c@{}} N + O\cite{tanmoy:2020:ec2} \end{tabular}}}    & \rot{{\begin{tabular}[c]{@{}c@{}} N+O+H \end{tabular}}} 
& \rot{{\begin{tabular}[c]{@{}c@{}}N \cite{Bartos:2016:invariantrep} \end{tabular}}}
       
        & \rot{{\begin{tabular}[c]{@{}c@{}}O \cite{das:2016:sematicsonline} \end{tabular}}}    & \rot{{\begin{tabular}[c]{@{}c@{}}H\cite{sayadi2018ensemble} \end{tabular}}}  & \rot{{\begin{tabular}[c]{@{}c@{}} N+O\cite{tanmoy:2020:ec2} \end{tabular}}}   & \rot{{\begin{tabular}[c]{@{}c@{}} N+O+H \end{tabular}}} 
   
        \\ \hline
    
         \hline
 Cryptominer & 0.80  & 0.87 & 0.93 & 0.87 &  0.82  & 0.2  & 0.01 & 0.14 & 0.11 & 0.013\\
         \hline
         Banker & 0.85  & 0.83 & 0.76 & 0.83 & 1 & 0.15  & 0.49 & 0.23  & 0.16   & \\
         \hline
         Spyware & 0.89 & 0.87 & 0.81 & 0.86  & 1 & 0.12  & 0.35 & 0.13  & 0.15 &  0   \\
         \hline
         Backdoor & 0.82 & 0.83 & 0.79 &  0.70 & 1 & 0.11  & 0.32 & 0.13 & 0.28 &0\\
         \hline
         Ransomware & 0.78 & 0.64 & 0.75 & 0.73 &  1 & 0.25  & 0.04 & 0.27 & 0.35 &  0\\
         \hline
         PUA & 0.83 & 0.72 & 0.74 & 0.81 &  0.99 & 0.12  & 0.35 & 0.28 &0.2 &  0.003\\
         \hline
         Downloader & 0.96 & 0.88 & 0.84  & 0.91 &  0.89 & 0.04  & 0.21 & 0.11 & 0.07 & 0.055\\
         \hline
         Deceptor & 0.87 & 0.88 & 0.78 &  0.86  & 0.78 & 0.13  & 0.51 & 0.24  & 0.15 &  0.051\\ \hline
         
         Mean & 0.85 & 0.82& 0.8 &0.82 & \cellcolor{green!20}\textbf{0.935} & 0.15  & 0.31 & 0.19 & 0.19 & \cellcolor{green!20}\textbf{0.015} \\
        \hline

    \end{tabular}
    {\scriptsize \bf *SIE- Same Input Ensembles, N - Network, O - OS, H - Hardware}
            \vspace{-0.4cm}

\end{table}

We next compare the resilience of \name using the case of cryptominer, which is best detected in hardware. Figure~\ref{fig:f1score_vs_load} plots the F1-Score of the cryptominer-specialized predictors based on network, OS, and hardware and \name on cryptominer data collected under varying system load conditions. As evident, the performance of hardware-based predictor though higher than OS and network at lower system loads, decreases significantly as load increases. In contrast, the OS-based predictor, agnostic to system load, outperforms both the network and hardware-based predictors as soon as more than $10$ additional user applications start executing simultaneously. Hence, the OS-based predictor is the most resilient to noise. In contrast, \name leverages the best of three worlds to achieve accurate and resilient malware detection (claims C-1 and C-2 in Section~\ref{sec:motivation}). At lower system loads, \name prioritizes network and hardware components for higher accuracy, whereas, at higher system loads, it uses the reliable OS component for prediction.

{\flushleft \bf Comparison with prior art.} Table~\ref{sundew:tab:priorartf1score} compares \name against our implementation of prior state-of-the-art predictors including single classifiers that rely on a single data source (network~\cite{Bartos:2016:invariantrep}, operating system~\cite{das:2016:sematicsonline}, or hardware~\cite{sayadi2018ensemble}); and, same-input ensembles that do not employ class-wise specialization~\cite{tanmoy:2020:ec2}. We compare these works based on the detection F1-Score and false-positive rate observed on the RaDaR dataset (Table~\ref{tab:dataset}). The cross-dimensional view of malware activity, specialization, and insightful aggregation of predictions in \name improve the detection F1-Score by at least 10\% as compared to these prior works (Table~\ref{sundew:tab:priorartf1score}A). Similarly, the class-specific specialization in \name decreases the false-positive by at-least 89\% as compared to the prior works (Table~\ref{sundew:tab:priorartf1score}B). Finally, with the incorporation of different data sources, we observe that \name is as resilient as the state-of-the-art OS-based works, even under noisy conditions.

\begin{figure}[t]
 \footnotesize
    \centering
  \includegraphics[width=\linewidth]{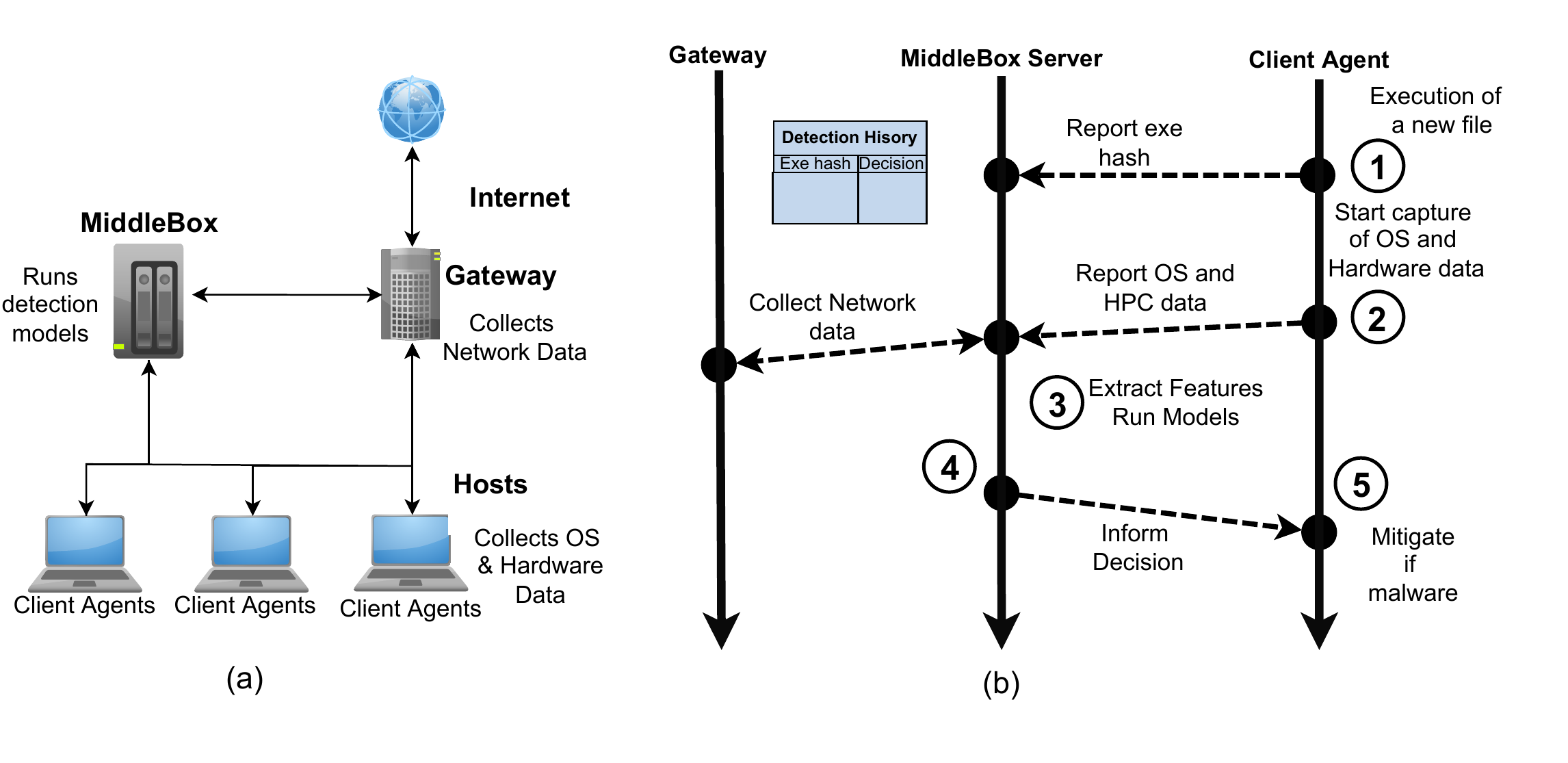} \vspace{-0.9cm}
  \caption {\footnotesize (a) An example deployment in an enterprise network. The network data is collected at the gateway, whereas the OS and hardware data are collected at the host machines. The middlebox server runs the \name framework. (b) The workflow of \nameA. .
  }
    \label{sundew:fig:design}
    \vspace{-0.5cm}
\end{figure}

{\flushleft \bf Overheads.} We next evaluate the overheads of \name considering an example deployment in an enterprise network. To measure the overheads, we first present the design and workflow of \name in the deployment in Figure~\ref{sundew:fig:design}. \name runs as a service on a middle-box server in the enterprise network, whereas the host machines run the client agents that enable the hosts to access the service to test any program. While the client agents collect the OS and hardware trails of the program under test, the gateway in the network collects the network behavior. Figure~\ref{sundew:fig:design}b illustrates the workflow of \name when a host accesses its service. At the host machine, whenever a new program executes, the client agent reports the hash of the program to the server (Step 1 in Figure~\ref{sundew:fig:design}b). To avoid repeated testing of the same program, \name  maintains a detection history of program hashes at the server. Thus, if the program is not analyzed previously, \name starts the capture of respective logs at the host, as well as the gateway for $2$ minutes (Step 2). After the duration, the server collects the OS and hardware data from the host, and the network data from the gateway (Step 3). The server then extracts the features from each data source, invokes the respective components of \name to predict the class of the program, and informs the final prediction to the host (Step 4). Finally, the client agent at the host takes the necessary action based on the prediction. 

We use GeekBench~\cite{geekbench} tool and observe that end-hosts incur an average overhead of $1.5\%$ at the first execution of a test program. Note that the end-hosts (client-agents) are responsible only for the collection of OS and hardware data, whereas the heavy-weight operations of feature extraction and specialized predictors run on the middle-box server (refer to Figure~\ref{sundew:fig:design}a).  While the middle-box server would require a dedicated provision of resources, the impact on users is minimal ($ 1.5\%$) and is restricted to the execution of new applications alone.

\vspace{-0.4cm}
\section{Discussion}
\label{sec:discussion}

In this section, we first discuss the applications of \nameA. Next, we discuss its limitations and present plausible directions for future work.
\vspace{-0.1cm}

{\flushleft \bf Applications.} The multi-featured approach and aggregation in \name can serve as an {\em analysis framework} for anti-virus companies and {\em defense solutions} for securing enterprises. As an analysis framework, the holistic view and specialization enable precise characterization of samples, thus reducing the manual efforts to label thousands of newer samples reported daily. Alternatively, \name can serve as defense solutions in enterprise networks to provide accurate and resilient detection of malware attacks.

\vspace{-0.1cm}
{\flushleft \bf Incremental update of predictors.} 
The \name ensemble involves predictors specialized for a set of malware classes. Further, aggregator functions are customized based on statistics from these predictors. With malware behavior evolving, the specialized predictors and aggregators would require updates. While mechanisms for incremental updates need to be explored in the future, we propose an auto-configuration engine that auto-configures the \name ensemble for any update or any deployment setting. Such an engine takes as inputs the labeled data from the three data sources and the user requirements per malware class. It outputs the ensemble, including its specialized predictors and aggregator functions. 

\vspace{-0.1cm}
{\flushleft \bf Scalability.} With a rampant increase in newly reported malware classes, the number of specialized predictors is bound to increase 3x (one for each data source), increasing the complexity of aggregator functions and overheads. Hence, specialized predictors for each class can get infeasible. A viable solution is to club models that share common features and user requirements in the 3-tuple to reduce the number of specialized predictors for each data source. We intend to build an automated framework to configure \name with an optimal number of specialized predictors in future work. Alternatively, Locality Sensitive Hashing (LSH) can assist in identifying the similarity of test programs to previously tested program hashes. Accordingly, LSH can assist in enabling only the relevant specialized predictor or data components to decrease overheads.

\vspace{-0.1cm}
{\flushleft \bf Extensive Characterization of Noise.} We analyze the impact of noise on \name using well-known benign applications. However, an extensive characterization of varying system load conditions and impact on the three data sources is planned for future work.

\vspace{-0.35cm}
\section{Conclusion}
\vspace{-0.2cm}
In this paper, we emphasize that malware classes are inherently different, and catering to the differences can improve the efficiency and resilience of detection. We propose \nameA, a novel multi-input ensemble of predictors and aggregator functions that leverages a multi-dimensional view of malware execution, considering its activities at the network, OS, and hardware and the system noise to provide a case-sensitive prediction. Our evaluations of \name on a real-world dataset indicate that the multi-dimensional view and specialization enable \name to avert infiltrating noise into the behavioral data while improving the accuracy, resilience, and false-positive guarantees. To the best of our knowledge, \name is the first to provide a multi-dimensional case-sensitive characterization of malware. The holistic approach and aggregation strategies open new avenues for malware research and detection models.
\label{sec:conclusion}

\ifCLASSOPTIONcaptionsoff
  \newpage
\fi

\bibliographystyle{ieeetr}
\renewenvironment{IEEEbiography}[1]
  {\IEEEbiographynophoto{#1}}
  {\endIEEEbiographynophoto}
\bibliography{IEEEabrv,main}

\vspace{-1.3cm}
\begin{IEEEbiography} {Sareena Karapoola}{\,}   is a Ph.D. scholar at the Department of Computer Science and Engineering, Indian Institute of Technology, Madras. Her research interests include the cyber-security, malware analysis and detection, machine learning for security, development of novel attack mitigation strategies, and testbeds for security research.\vspace{-1.5cm}
\end{IEEEbiography}

\begin{IEEEbiography}{Nikhilesh Singh}{\,} is a Ph.D. student at Department of Computer Science and Engineering, Indian Institute of Technology, Madras. His research interests include the deployment of Machine Learning for safety and system security including malware defenses, micro-architectural security, operating system security, secure hardware designs. \vspace{-1.5cm}
\end{IEEEbiography}

\begin{IEEEbiography}{Chester Rebeiro}{\,} is an Associate Professor at the Department of Computer Science and Engineering, Indian Institute of Technology, Madras. His research interests include hardware security, operating systems security, and applied cryptography. He also leads  the  effort  at  designing  secure  RISC-V  micro-processors  for  embedded  platforms at IIT Madras.\vspace{-1.5cm}
\end{IEEEbiography}

\begin{IEEEbiography}{Kamakoti V.} is a Professor and the Director of IIT Madras. He specializes in the areas of computer architecture, secure systems engineering, and network security and privacy. He is a coordinator of the Information Security Education and
Awareness program of the Department of Information Technology, Government of India and the Chairman of the Task Force on Artificial Intelligence for India’s Economic Transformation. He has also won several awards such as the IBM Faculty Award (2016).
\end{IEEEbiography}
\end{document}